\documentclass[11pt,reqno,twoside]{amsart}
\usepackage[latin1]{inputenc} 
\usepackage{amssymb,amsfonts,amsmath}
\usepackage[english]{babel}
\usepackage{pstricks}
\usepackage{longtable}
\usepackage{graphicx, subfigure}
\usepackage[foot]{amsaddr}
\usepackage{array}
\usepackage{color}
\usepackage{enumerate}

\newtheorem{Proposition}{Proposition}
\newtheorem{Definition}{Definition}

\begin{document}

\title[A hybrid mathematical model for the zebrafish lateral line]{A hybrid mathematical model for self-organizing cell migration in the zebrafish lateral line}

\author[E. Di Costanzo]{Ezio Di Costanzo$^1$}
\address{$^1$Dipartimento di Scienze di Base e Applicate per l'Ingegneria -- Sezione di Matematica, Sapienza University of Rome, Via A. Scarpa 16, 00161 Rome, Italy.}
\author[R. Natalini]{Roberto Natalini$^2$}
\address{$^2$Istituto per le Applicazioni del Calcolo ``M. Picone'' -- Consiglio Nazionale delle Ricerche, Via dei Taurini 19, 00185 Rome, Italy.}

\author[L. Preziosi]{Luigi Preziosi$^3$}
\address{$^3$Department of Mathematical Sciences, Politecnico di Torino, Corso Duca degli Abruzzi 24, 10129 Turin, Italy.}

\email{$^1$ezio.dicostanzo@sbai.uniroma1.it}
\email{$^2$roberto.natalini@cnr.it}
\email{$^3$luigi.preziosi@polito.it}


\begin{abstract}
In this paper we propose a \emph{discrete in continuous} mathematical model for the morphogenesis of the posterior lateral line system in zebrafishes. Our model follows closely the results obtained in recent biological experiments. We rely on a hybrid description: discrete for the cellular level and continuous for the molecular level. We prove the existence of steady solutions consistent with the formation of particular biological structure, the neuromasts. Dynamical numerical simulations are performed to show the behavior of the model and its qualitative and quantitative accuracy to describe the evolution of the cell aggregate.
\end{abstract}
\maketitle
\numberwithin{equation}{section}
\numberwithin{figure}{section}
\section{Introduction}
The \emph{lateral line} is a sensory system, which is present in fish and amphibians, that is used to detect movement and vibration in the surrounding water and is involved in a large variety of behaviors, from prey detection to predator avoidance, school swimming and sexual courtship. It extends from the head to the tail along each flank of the fish,  and it is formed by a set of sensory organs, the \emph{neuromasts}, arranged on the body surface in specific patterns. The neuromasts, located between the ear and the eye, form the so-called anterior lateral line system (ALL), while neuromasts on the body and tail form the \emph{posterior lateral line} system (PLL) \cite{ghysen, coombs}.
\par In this paper we propose and analyze a mathematical model for the morphogenesis of the \emph{zebrafish} (\emph{Danio rerio}) PLL primordium. The development of this sensory organ represents a subject of general importance, as a paradigm to understand the growth, regeneration, and self-organization of other organ systems during development and disease \cite{chitnis}. Recent studies \cite{gilmour1, nechiporuk, gilmour} (see also \cite{haddon, itoh, draper, ghysen, li, matsuda, sarrazin, mizoguchi, sweet}) have investigated migration and self-organization in the zebrafish lateral line system, where a complex system of receptor activation drives embryonic cells, rather than a guidance determined by birth. However, the complete mechanism for cells arranging and organization is still relatively poorly understood \cite{gilmour}.
\par
\begin{figure}[htbp]
	\centering
\subfigure[]{\includegraphics[width=0.7\textwidth]{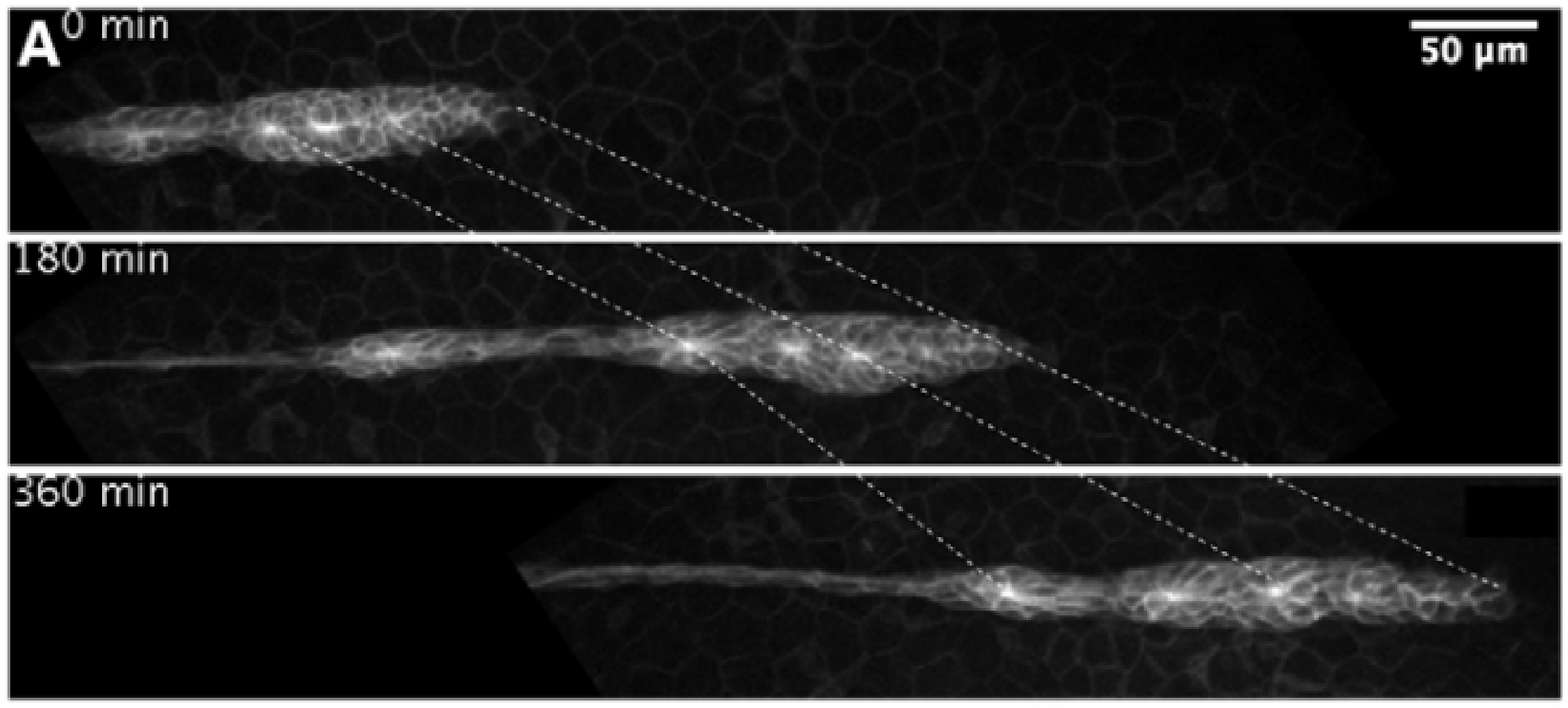}}\\
\vspace{0cm}
\subfigure[]{\includegraphics[width=0.9\textwidth]{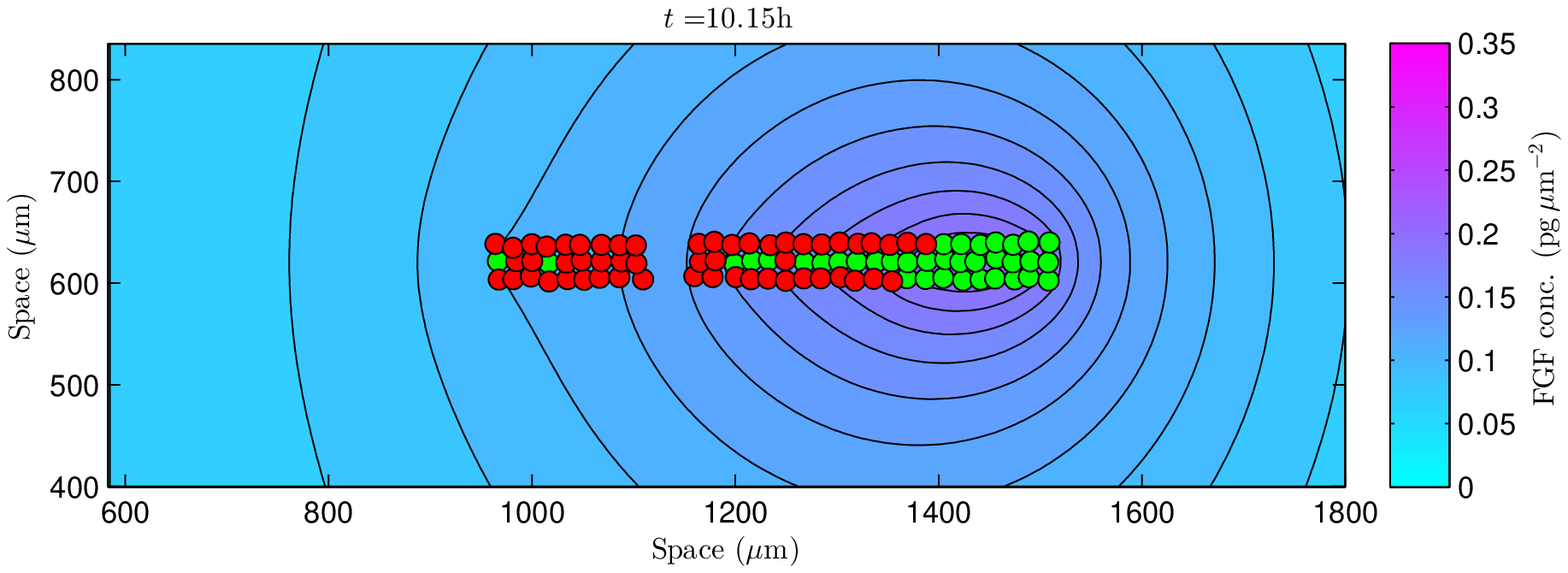}}
\caption{(a) Images from a time-lapse movie about the migration in the zebrafish PLL primordium. Leading zone is on the right of the primordium, trailing zone is on the left. Cell migration is to the right while neuromasts deposition occurs in the trailing region (source reproduced with permission from \cite{gilmour}). (b) An example of dynamical simulation of the our mathematical model (see Section \ref{sec:dynamic} for further details).}
\label{fig:gilmourZF}
\end{figure}

 Loosely speaking, lateral line formation consists in a group of mesenchymal cells that migrate driven by a haptotactic signal. In a second phase, a process of differentiation in the rear of the migrating group induces a mesenchymal-epithelial transition that is at the origin of the detachment of rosette-shaped structures. This corresponds to the growth and location of the neuromasts along the two flanks of the primordium (see Figure \ref{fig:gilmourZF} (a) below, from \cite{gilmour}).

\par Our aim is to obtain a minimal mathematical model which is able to:
\begin{enumerate}[a)]
	\item describe the collective cell migration, the formation and the detachment of the neuromasts, in the spatial and temporal scale of the experimental observations;
\item ensure the existence and stability of the rosette structures of the emerging neuromasts, as stationary solutions of the system.
\end{enumerate}
Request b), among other, allow us to obtain some important restrictions on the range of the values of some parameters of the model, which will be used  in the numerical simulations of the dynamical case, other parameters being obtained from biological literature or by a numerical fitting.
\par The model proposed here is a hybrid model: it describes cells as discrete entities and chemotactic molecules as continuous concentrations. This is a reasonable choice if we think that the total number of cells involved in the morphogenesis process is in the range 80--100 \cite{gilmour1, sarrazin}. For analytical and computational simplicity in our analysis, here we consider only the  2D case, although we do not expect great changes passing to 3D, since experimental observations suggest that these phenomena involve only a thin cell layer. As we will see in more detail through numerical simulations given in Section \ref{sec:dynamic}, we can state that our mathematical model shows a substantial agreement with the biological observations and with the experimental data proposed in literature (Figure \ref{fig:gilmourZF} (b), to be compared with (a)). 
\par From a mathematical point of view, our model is based on a second order equation of the form
\begin{align*}
	\ddot{\mathbf{X}}_i=\mathbf{F}(t,\mathbf{X},\dot{\mathbf{X}},u,\nabla u)-\mu\dot{\mathbf{X}}_i,
\end{align*}
where $\mathbf{X}_i$, $i=1,\dots,N_{\text{tot}}$, is the position vector of the $i$-th cell, $N_{\text{tot}}$ is the total number of cells, $\mathbf{X}:=(\mathbf{X}_1,\dots,\mathbf{X}_{N_{\text{tot}}})$, and $\dot{\mathbf{X}}:=(\dot{\mathbf{X}}_1,\dots,\dot{\mathbf{X}}_{N_{\text{tot}}})$. The function $\mathbf{F}$ includes several effects: from the detection of chemical signals $u$ (chemotaxis, lateral inhibition) to mutual interactions between cells (alignment, adhesion, repulsion). All these effects take into account a non local sensing radius. In particular, we included an alignment term inspired by the Cucker-Smale mechanism \cite{cucker}, though in our case it is coupled with other effects. The term $-\mu\dot{\mathbf{X}}_i$ represents damping due to cell adhesion to the substrate. Chemical signals are described using a reaction-diffusion equation
\begin{align*}
	\partial_t u=D\Delta u+S(t,\mathbf{X},u),
\end{align*}
with a possible source or degradation term given in $S$. Finally the cell mesenchymal-epithelial differentiation in the primordium is performed by a switch variable, whose evolution in time is given by a suitable threshold function.
\par We will show that our model admits particular stationary solutions, biologically relevant and consistent with experimental observations. They correspond to the so called rosettes, that will form the future neuromasts. We investigate numerically their stability, finding in turn a nice agreement with biological evidences both in the stationary and in the dynamical setting. 
\par The paper is organized as follows: in Section \ref{sec:bio} we recall some biological backgrounds and the phenomenology, based on the existing experimental observations. In Section \ref{sec:math:model} the mathematical model is introduced and its main features are discussed. In Section \ref{sec:steady} we study the stationary configuration of the rosettes, and their stability. Section \ref{sec:dynamic} deals with the dynamic model. We explain the methods used in the numerical simulations and some 2D numerical tests are presented, with the aim of illustrating the power and the limits of our approach. Section \ref{sec:conclusion} is devoted to the conclusions. Finally,  Appendix \ref{ap:parameter} contains the lists of main dimensional and nondimensional parameters used in the model.
\section{Biological backgrounds}\label{sec:bio}
About zebrafish PLL, recent studies and experimental observations \cite{gilmour1, gilmour, nechiporuk} show an initial elongated single group of cells in the otic vesicle, in which we can distinguish a \emph{trailing region} near the head and a \emph{leading region} oriented towards the future tail of the embryo. In the following few hours after the fertilization, a total cell migration begins posteriorly, from head to tail. Then the cells in the trailing region assemble into rosette structures (proto-neuromast), that are progressively deposed during the migration to form \emph{neuromasts} \cite{nechiporuk} (see again Figure \ref{fig:gilmourZF} (a)).
\par In general we can state two primary mechanisms that concur to this morphogenesis process: a collective migration, and the neuromasts assembly.
About cell migration a very important role is played by chemokines as the \emph{stromal cell-derived factor-1a} (\emph{SDF-1a}) and its \emph{CXCR4b} receptor \cite{ghysen, gilmour1}. The former is expressed by the horizontal myoseptum, that separates the dorsal and ventral axial muscles, and acts as a haptotactic stripe for the migrating cells; the latter is expressed by the primordium itself \cite{li}. Chemokine signaling is necessary to drive migration. In fact, it has been proved that, in absence of CXCR4b, cell movements are strongly uncoordinated, with a ``zig zag'' pattern, as can be seen performing kymograph analysis (see \cite{gilmour1}). Moreover, next to the chemoattractant chemokine, migration within the primordium is guaranteed by a \emph{cell-cell interaction}, exerted by an adhesion force via filopodia. This is confirmed by two observations in \cite{gilmour1}: first, cells lacking CXCR4b receptor, transplanted into wild-type primordium, preserve their migration through the contact with neighboring cells; secondly, even a small number of wild-type cells, transplanted in a mutant primordium (lacking the SDF-1a receptor), after moving themselves toward the leading edge can restore the collective migration. In this contest other studies \cite{liu2, kerstetter, papusheva, liu, mertz} have shown that intercellular adhesion, typically through molecules as the cadherins, and \emph{cell-substratum adhesion}, through integrins, have a crucial role in the spatial organization of tissues and in embryonic development. Integrin- and cadherin-mediated adhesion allows cells and tissues to respond to mechanical stimuli from their environment and to change shape without loosing integrity (\cite{papusheva} and references therein).
\par To understand the mechanism which drives the rosettes organization and neuromast deposition (a mechanism however not yet completely described \cite{gilmour}), we have to make some considerations. Neuromast formation is strongly influences by the concentration of \emph{fibroblast growth factors} (\emph{FGFs}) and their receptors \emph{FGFRs} \cite{nechiporuk, gilmour}. In zebrafishes,  FGF signaling drives cells to assemble rosettes and gives rise to the subdivision of the lateral line. Among the 22 members of the FGF family, only FGF3 and FGF10 are expressed by the primordium \cite{bottcher}, and they are substantially equivalent. In fact, inactivation of FGF3 or FGF10 alone does not alter significantly the development of the primordium, demonstrating a robustness of the system \cite{gilmour}. On the other hand, using a FGFR inhibitor SU5402, strongly affects the primordium: cells became disorganized and neuromast deposition stops. In this case also the collective migration is compromised, probably due to an alteration also in CXCR4b receptors \cite{nechiporuk}. Therefore, the rosette formation depends mainly on the total level of FGF and, as we will see below, on its location. We observe also that FGF and FGFR expressions are mutually exclusive, as confirmed by the location of their molecules: the former, broadly expressed in the leading region, and focused in one or two cells at the center of the rosettes in the trailing region, the latter, at the same time, expressed in the trailing region except the FGF foci \cite{nechiporuk, gilmour}. In the aim of designing a mathematical model, this suggests to divide the cell population into two groups: the \emph{leader} mesenchymal cells (expressing FGF), and the \emph{follower} epithelial cells (expressing FGFR). At the beginning, all the cells belong to the leading group. Afterwards, few minutes after fertilization, some leaders in the trailing zone start to become followers, except one or two leader cells located in the center, that maintain their mesenchymal state. Loss of FGF activity, on the contrary, implies that no transition can occurs \cite{gilmour}. Then, the follower cells are driven towards the FGF source to form a rosette. As a proto-neuromast becomes fully mature, it is deposited from the trailing edge, and a new rosette is formed again in a cyclic mechanism.
\par Let us now  propose some rules that model the transition from a leader cell into a follower, corresponding to the activation of the FGFR receptor. We assume in the following that the transition occurs under  three concomitant  conditions:
\begin{enumerate}
	\item a low level of SDF-1a \cite{ghysen};
	\item a  high level of FGF \cite{gilmour};
	\item a low number of  followers in the neighborhood.
\end{enumerate}
\par The first condition implies that rosette formation begins in the trailing region, where SDF-1a signal is already degraded by cells in the leading edge. The third condition translates a common phenomenon in embryology, the so called \emph{lateral inhibition}: a cell that adopts a particular feature inhibits its immediate neighbours from doing likewise. This condition, together with the second, implies that followers, activating FGFR receptor, inhibit the same activation in a more surrounded central leader, so that it will express a significantly higher level of FGF signal \cite{haddon, itoh, hart, matsuda, mizoguchi, sweet}. Finally we remark that the leader-follower transition can be reversible, possibly with a time delay. In fact, blocking FGF activity makes all cells equally leader and causes the consequent melting of formed rosettes \cite{gilmour}.
\par In the following we will see that our mathematical model will be consistent with the biological observations if we consider chemical effects concomitant with other cell mechanisms, as lateral inhibition, alignment, and adhesion-repulsion effects.
%
%
%
%AGGIUNGERE:schema trailing e leading region, schema interazioni segnale, 
%
%
%
%
\section{The mathematical model}\label{sec:math:model}
According to the above observations we propose a hybrid model which takes into account the difference between the cellular and the chemical scale. At the cellular level the model is discrete and includes the equation of the motion and the equation of state leader-follower for each single cell, while at the molecular level the model is continuous and is based on the equations for the various chemical signals involved. Let us summarize the main ingredients which compose our model. 
\par For the cell motion we use a second order dynamic equation, which takes into account the forces acting on the cells. These forces are given by chemical signals and mechanical interaction between cells. Because of the equivalence between FGF3 and FGF10, we consider a single concentration and a single receptor, that we will denote respectively as FGF and FGFR. The SDF-1a effect is described by a haptotactic term produced by the gradient of the concentration of this chemokine, see  \cite{eisenbach} for some biological backgrounds, while mathematical references can be found in \cite{murrayII, perthame}. In the same way, the action of FGFR on a follower cell is described by a chemotactic effect due to the gradient of the FGF produced by a leader cell. 

\par Next we describe the cell-cell mechanical interactions due to filopodia, which consist in an alignment effect and both a radial attraction and repulsion depending on the relative position of the cells, see  \cite{mertz} for experimental results in this direction. About alignment effects, we base our description on the seminal paper  by F. Cucker and S. Smale \cite{cucker}, while for the attraction-repulsion effects we refer to the mechanism introduced by  D'Orsogna et al. \cite{dorsogna}; both effects are considered by \cite{albi}. Finally, we introduce a damping term, proportional to the velocities, which models  cell adhesion to the substrate \cite{rubinstein, fournier, bayly}.
\par The follower-leader differentiation is represented mathematically by a switch state variable, which change its value according to the level of some related functions, which take into account  the concentration of SDF-1a and of FGF and also the number of cells in a given neighborhood of the given cell.       
\par About the concentration of the FGF signal we associate a diffusion equation including a source term, given by the FGF production of the leader cells, and a natural molecular degradation term. The concentration of SDF-1a is described by an equation involving its degradation during the haptotactic process. 
\subsection{The basic mathematical model}
Starting from the above considerations,  we start to work in the  following framework:
\begin{center}
\fbox{
\begin{minipage}{0.9 \textwidth}
\begin{align*}
\begin{array}{l}
\text{acceleration of}\\
\text{$i$-th cell}
\end{array}	
	&=\text{haptotactic effect of SDF-1a}\\
	&+\text{chemotactic effect of FGF source on the followers}\\
	&+\text{cells alignment}+\text{cell adhesion and repulsion}+\text{damping effect}\\
\text{$i$-th cell kind}&=
\left\{
\begin{array}{lll}
\text{follower}&\text{if}& \text{low level of SDF-1a}+\text{hight level of FGF}\\ 	  		 
&&+\text{absence of lateral inhibition}\\
\text{leader}&&\text{otherwise} 
\end{array}	
\right.\\
\begin{array}{l}
\text{rate of change of}\\
\text{FGF signal}
	\end{array}	
	&=\text{diffusion}+\text{production}+\text{molecular degradation}\\
\begin{array}{l}
\text{rate of change of}\\
\text{SDF-1a signal}
	\end{array}	
	&=\text{degradation}
\end{align*}
\end{minipage}
}
\end{center}
\medskip
Let $\mathbf{X}_i(t)$ be the position of a single $i$-th cell, $s(\mathbf{x},t)$ the SDF-1a concentration, $f(\mathbf{x},t)$ the total FGF concentration (including both FGF3 and FGF10), $\varphi_i(t)$ a variable that distinguishes a $i$-th cell to be, at time $t$, a leader ($\varphi_i(t)=1$) or a follower ($\varphi_i(t)=0$). We introduce the following equations:
\begin{equation}\label{sysF}
\left\{
\begin{array}{rl}
\ddot{\mathbf{X}}_i&=\alpha \mathbf{F}_1\left(\nabla s\right)+\gamma(1-\varphi_i)\mathbf{F}_1\left(\nabla f\right)+\mathbf{F}_2(\dot{\mathbf{X}})+\mathbf{F}_3(\mathbf{X})-\left[\mu_\text{F}+(\mu_\text{L}-\mu_\text{F})\varphi_i\right]\dot{\mathbf{X}}_i,\\\\
\varphi_i&=
\left\{
\begin{array}{lll}
0,&\text{if}& \delta F_1(s)-\left[k_\text{F}+(k_\text{L}-k_\text{F})\varphi_i\right]F_1(h(f))+\lambda\Gamma(n_i)\leq0,\\
1,&&\text{otherwise},
\end{array}
\right.\\\\
\partial_t f&=D\Delta f+\xi F_4(\mathbf{X})-\eta f,\\\\
\partial_t s&=-\sigma s F_5(\mathbf{X}),
\end{array}
\right.
\end{equation}
where $\alpha$, $\gamma$, $\mu_\text{L}$, $\mu_\text{F}$, $\delta$, $k_\text{L}$, $k_\text{F}$, $\lambda$, $D$, $\xi$, $\eta$, $\sigma$ are given positive constants, and $F_n(\cdot)$, $n=1,\dots,5$, are suitable functions. 
\par The term $F_1$, which is related to the detection of a chemical signal by $i$-th cell in its neighborhood, is taken to be a weighted average over a ball of radius $\bar{R}$ and centered in $\mathbf{X}_i$:
\begin{align}\label{F_chemicals}
	F_1(g(\mathbf{x},t)):=\frac{1}{W}\int_{\mathbf{B}(\mathbf{X}_i,\bar{R})}g(\mathbf{x},t)w_i(\mathbf{x})\,d\mathbf{x},
\end{align}
where 
\begin{align}
\mathbf{B}(\mathbf{X}_i,\bar{R}):=\left\{\mathbf{x}:\left|\left|\mathbf{x}-\mathbf{X}_i\right|\right|\leq \bar{R}\right\},
\end{align}
$\left|\left|\cdot\right|\right|$ being the Euclidean norm, 
\begin{align}
w_i(\mathbf{x}):=
\left\{
\begin{array}{lll}
2\exp\left(-||\mathbf{x}-\mathbf{X}_i||^2\displaystyle\frac{\log 2}{\bar{R}^2}\right)-1,&\text{if}&||\mathbf{x}-\mathbf{X}_i||\leq\bar{R};\\
0,&&\text{otherwise};
\end{array}
\right.
\end{align}
is a truncated Gaussian weight function, and
\begin{align}
	W:=\int_{\mathbf{B}(\mathbf{X}_i,\bar{R})}w_i(\mathbf{x})\,d\mathbf{x},
\end{align}
independently of $i$. A similar definition holds for the vector quantity $\mathbf{F}_1$. Reasonably we will choose $\bar{R}$ larger than the cell radius $R$ (see Appendix \ref{ap:parameter}), so (\ref{F_chemicals}) describes a chemical signal that is sensed more in the center of the cell and less at the edge of the cell extensions. The second addend in (\ref{sysF})$_1$ refers to the attraction of a follower cell toward a source of FGF ligand. The switch variable $\varphi_i$ makes this term zero for a leader cell that, expressing FGF, does not activate FGFR receptor (see Section \ref{sec:bio}). 
\par The effect included in the third addend of (\ref{sysF})$_1$ represents a possible cell alignment. For it we assume a Cucker-Smale \emph{flocking term}:
\begin{align}\label{alignment-new}
	\mathbf{F}_2(\dot{\mathbf{X}}):=\frac{1}{\bar{N}_i}\sum_{j:\mathbf{X}_j\in \mathbf{B}(\mathbf{X}_i,R_1)\backslash\left\{\mathbf{X}_i\right\}}\mathbf{H}(\dot{\mathbf{X}}_j-\dot{\mathbf{X}}_i).
\end{align}
Here $R_1$ is a suitable radius of influence,
\begin{align}\label{alignment-new2}
	\bar{N}_i:=\text{card}\left\{j:\mathbf{X}_j\in \mathbf{B}(\mathbf{X}_i,R_1)\backslash\left\{\mathbf{X}_i\right\}\right\},
\end{align}
and the function $\mathbf{H}$ depends on the relative velocities $\dot{\mathbf{X}}_j-\dot{\mathbf{X}}_i$, i.e.:
\begin{align}\label{alignment2}
\mathbf{H}(\dot{\mathbf{X}}_j-\dot{\mathbf{X}}_i):=\left[\beta_\text{F}+(\beta_\text{L}-\beta_\text{F})\varphi_i\varphi_j\right]\frac{R_1^2}{R_1^2+||\mathbf{X}_j-\mathbf{X}_i||^2}(\dot{\mathbf{X}}_j-\dot{\mathbf{X}}_i),
\end{align}
$\beta_\text{F}$, $\beta_\text{L}$ being constants. In particular we can have different coefficient of alignment for a leader or follower cell: the product $\varphi_i\varphi_j$ makes the coefficient equal to $\beta_\text{F}$ if at least one of the two cell is follower ($\varphi_i\varphi_j=0$) and equal to $\beta_\text{L}$ in the case of two leaders ($\varphi_i\varphi_j=1$). We remark that the flocking term given by (\ref{alignment2}), which is studied in \cite{cucker, ha}, in our model is coupled with other effects, as chemotaxis and attraction-repulsion effects (see below), and it is also computed on a truncated domain. In the following we will assume $R_1=\bar{R}$ (see Appendix \ref{ap:parameter}), but in principle they can be different.
\par Function $\mathbf{F}_3$ includes adhesion-repulsion effects. In particular repulsion occurs at a distance between the centers of two cells less than $R_4$ and takes into account the effects of a possible cell deformation. Conversely, adhesion occurs at a distance greater than $R_4$ and less than $R_5>R_4$, and it is due to a mechanical interaction between cells via filopodia. We assume
\begin{align}
\mathbf{F}_3(\mathbf{X}):=\sum_{j:\mathbf{X}_j\in\mathbf{B}(\mathbf{X}_i,R_5)\backslash\left\{\mathbf{X}_i\right\}}\mathbf{K}(\mathbf{X}_j-\mathbf{X}_i),
\end{align}
where the function $\mathbf{K}$ depends on the relative positions $\mathbf{X}_j-\mathbf{X}_i$, i.e.:
\footnotesize
\begin{align}\label{kappa}
\mathbf{K}(\mathbf{X}_j-\mathbf{X}_i):=
\left\{
\begin{array}{lll}
-\omega_{\text{rep}}\left(\displaystyle\frac{1}{||\mathbf{X}_j-\mathbf{X}_i||}-\frac{1}{R_4}\right)\displaystyle\frac{\mathbf{X}_j-\mathbf{X}_i}{||\mathbf{X}_j-\mathbf{X}_i||},&\text{if}& ||\mathbf{X}_j-\mathbf{X}_i||\leq R_4;\\\\
\left[\omega_{\text{adh,F}}+(\omega_{\text{adh,L}}-\omega_{\text{adh,F}})\varphi_i\varphi_j\right]\left(||\mathbf{X}_j-\mathbf{X}_i||-R_4\right)\displaystyle\frac{\mathbf{X}_j-\mathbf{X}_i}{||\mathbf{X}_j-\mathbf{X}_i||},&\text{if}& R_4<||\mathbf{X}_j-\mathbf{X}_i||\leq R_5;
\end{array}
\right.
\end{align}
\normalsize
$\omega_{\text{rep}}$, $\omega_{\text{adh,L}}$, $\omega_{\text{adh,F}}$ being constants. In practice we will choose $R_4=2R$ (see Appendix \ref{ap:parameter}), so that repulsion occurs when two cells start to be effectively overlapped. We note that function (\ref{kappa})$_1$ gives a repulsion which goes as $1/r$, $r$ being the distance between the centers of two cells, as we can find in \cite{cristiani-piccoli-tosin, scianna}. The function (\ref{kappa})$_2$ represents Hooke's law of elasticity, with different elastic coefficients for a leader cell and for a follower. In particular we have $\omega_{\text{adh,F}}$ if at least one of the two cells is a follower ($\varphi_i\varphi_j=0$) and $\omega_{\text{adh,L}}>\omega_{\text{adh,F}}$ if two cells are both leader ($\varphi_i\varphi_j=1$). Similar terms can be found in \cite{albi} and references therein. We remark that adhesion (\ref{kappa})$_2$ and alignment (\ref{alignment2}) produce different effects, though they both refer to a cell-cell interaction: the former a radial effect, the latter a tangential effect. 
\par The last term in the first equation is due to the cell adhesion to the substrate (see for example \cite{rubinstein, fournier, bayly}), possibly with a different damping coefficient for a leader ($\mu_\text{L}$, given by $\varphi_i=1$) or a follower cell ($\mu_\text{F}$, given by $\varphi_i=0$).
\par The second equation in (\ref{sysF}) defines the switch variable $\varphi_i$ for the $i$-th cell. The leader-to-follower transition is performed requiring that the threshold function at the right hand side of (\ref{sysF})$_2$ is less than zero, according with the three conditions described in Section \ref{sec:bio}. For the FGF detection in $F_1(h(f))$ we choose the following form for the function $h(f)$:
\begin{align}\label{fun-fgf-fgfmax}
h(f):=\frac{f}{f_{\max}+f},
\end{align}
where $f_{\max}$ is constant. The function (\ref{fun-fgf-fgfmax}) includes a possible saturation effect when FGF molecules tend to occupy all receptors of a cell. The coefficients $k_\text{L}$ and $k_\text{F}$, related respectively to a leader and a follower cell, provides a delay in the inverse follower-to-leader transition setting suitably $k_\text{L}<k_\text{F}$, this in accordance to observations in Section \ref{sec:bio}. About the lateral inhibition mechanism, we introduce a function $\Gamma(n_i)$ that counts the number $n_i$ of cells in a suitable neighborhood of the $i$-th cell, with radius of influence $R_2$, namely
\begin{align}\label{gamma-n-i}
	\Gamma(n_i):=\frac{e^{n_i}}{e^{n_i}+\Gamma_0}-\frac{1}{1+\Gamma_0},
\end{align}
where
\begin{align}\label{n-i}
	n_i:=\text{card}\left\{j:\mathbf{X}_j\in \mathring{\mathbf{B}}(\mathbf{X}_i,R_2)\backslash\left\{\mathbf{X}_i\right\}\right\},
\end{align}
$\Gamma_0$ is a constant, and $\mathring{\mathbf{B}}(\mathbf{X}_i,R_2)$ is the interior of $\mathbf{B}(\mathbf{X}_i,R_2)$. Function (\ref{gamma-n-i}) is justified if we think of a neuromast as made by a single central leader and some followers around. In this contest it makes an appropriate difference between the number of cells counted by the central cell and those counted by a lateral cell. Moreover, it provides a fast saturation effect when $n$ increases, so that a central leader counts about the same number of cells from a certain value of $n$. This is useful to describe the possibility to obtain neuromasts with a variable number of cells, according to experimental observations (generally 8-20 cells) \cite{gilmour}. A suitable value for the constant $\Gamma_0$ is given in Appendix \ref{ap:parameter}. 
\par In the diffusion equation (\ref{sysF})$_3$, only leader cells are responsible of the production of FGF, so that
\begin{align}\label{F7-eq}
F_4(\mathbf{X}):=\sum_{j=1}^{N_{\text{tot}}}\varphi_j\chi_{\mathbf{B}(\mathbf{X}_j,R_3)},
\end{align}
where $N_{\text{tot}}$ is the total number of cells, and
\begin{align}
\chi_{\mathbf{B}(\mathbf{X}_j,R_3)}:=
\left\{
\begin{array}{lll}
1,&\text{if}&\mathbf{x}\in\mathbf{B}(\mathbf{X}_j,R_3);\\
0,&&\text{otherwise}.
\end{array}
\right.
\end{align}
\par Similarly, in equation (\ref{sysF})$_4$  we take
\begin{align*}
F_5(\mathbf{X}):=\sum_{j=1}^{N_{\text{tot}}}\displaystyle\chi_{\mathbf{B}(\mathbf{X}_j,R_3)},
\end{align*}
in which the variable $\varphi$ does not appear now, because both leaders and followers are involved in the haptotactic process. Typically, we will choose $R_3=R$ considering that the source of FGF and the degradation of SDF-1a signal is substantially defined by the dimension of a single cell (see Appendix \ref{ap:parameter}).
\par The above observations let us to summarize the following model:
\footnotesize
\begin{equation}\label{sys-dim}
\begin{split}
&\left\{
\begin{array}{lll}
\ddot{\mathbf{X}}_i&=\displaystyle\frac{\alpha}{W}\int_{\mathbf{B}(\mathbf{X}_i,\bar{R})}\nabla s(\mathbf{x},t)w_i(\mathbf{x})\,d\mathbf{x}+\displaystyle	\frac{\gamma(1-\varphi_i)}{W}\int_{\mathbf{B}(\mathbf{X}_i,\bar{R})}\nabla f(\mathbf{x},t)w_i(\mathbf{x})\,d\mathbf{x}\\\\
	&+\displaystyle\frac{1}{\bar{N}_i}\sum_{j:\mathbf{X}_j\in \mathbf{B}(\mathbf{X}_i,R_1)\backslash\left\{\mathbf{X}_i\right\}}\mathbf{H}(\dot{\mathbf{X}}_j-\dot{\mathbf{X}}_i)+\sum_{j:\mathbf{X}_j\in \mathbf{B}(\mathbf{X}_i,R_5)\backslash\left\{\mathbf{X}_i\right\}}\mathbf{K}(\mathbf{X}_j-\mathbf{X}_i)-\left[\mu_\text{F}+(\mu_\text{L}-\mu_\text{F})\varphi_i\right]\dot{\mathbf{X}}_i,\\\\
\displaystyle\varphi_i&\displaystyle=\left\{
\begin{array}{lll}
0&\text{if}&
\displaystyle\frac{\delta}{W}\int_{\mathbf{B}(\mathbf{X}_i,\bar{R})}s(\mathbf{x},t)w_i(\mathbf{x})\,d\mathbf{x}-\frac{k_\text{F}+(k_\text{L}-k_\text{F})\varphi_i}{W}\int_{\mathbf{B}(\mathbf{X}_i,\bar{R})}\frac{f(\mathbf{x},t)}{f_{\text{max}}+f(\mathbf{x},t)}w_i(\mathbf{x})\,d\mathbf{x}\\\\
&&+\displaystyle\lambda \Gamma(n_i)\leq0,\\\\
1&&\text{otherwise},
\end{array}
\right.
\\\\\displaystyle	\partial_t f&\displaystyle=D\Delta f+\xi\sum_{j=1}^{N_{\text{tot}}}\displaystyle\varphi_j\chi_{\mathbf{B}(\mathbf{X}_j,R_3)}-\eta f,\\\\
\displaystyle	\partial_t s&\displaystyle=-\sigma s\sum_{j=1}^{N_{\text{tot}}}\displaystyle\chi_{\mathbf{B}(\mathbf{X}_j,R_3)},
	\end{array}
	\right.\\
	\end{split}
\end{equation}
\normalsize
where the functions $\mathbf{H}(\dot{\mathbf{X}}_j-\dot{\mathbf{X}}_i)$ and $\mathbf{K}(\mathbf{X}_j-\mathbf{X}_i)$ are given by (\ref{alignment2}) and (\ref{kappa}). Initial and boundary conditions have to be specified. For the $i$-th cell we set
\begin{align}\label{pos-vel-iniz}
\mathbf{X}_i(0)=\mathbf{X}_{i0};\quad\mbox{and}\quad\dot{\mathbf{X}}_i(0)=\mathbf{0},\quad i=1,\dots,N_{\text{tot}},
\end{align}
together with the equally initial cell state of leader:
\begin{align}\label{phi-iniz}
&\varphi_i(0)=1,\quad i=1,\dots,N_{\text{tot}}.
\end{align}
Now, let $\Omega=[a,b]\times[c,d]$ our domain, for FGF signal we require zero initial concentration and homogeneous Neumann boundary condition:	
\begin{align}\label{f-iniz-bound}
	f(\mathbf{x},0)=0;\quad \frac{\partial f}{\partial\mathbf{n}}=0,\quad\mbox{on $\partial\Omega$}.
\end{align}
No-flow boundary condition (\ref{f-iniz-bound})$_2$ is justified if we think of an experiment in which our domain is isolated from the surrounding environment. Then, since initially SDF-1a is only located in a given region
\begin{align}\label{s-iniz1}
s(\mathbf{x},0)=s_0(\mathbf{x}),
\end{align}
where $s_0(\mathbf{x})$ has compact support in $\Omega$. In particular we consider a rectangular stripe of width $2l$, $[\bar{a},\bar{b}]\times [m-l,m+l]$, with $[\bar{a},\bar{b}]\subset [a,b]$, $m=(c+d)/2$, and \begin{align}\label{s-iniz2}
	s_0(x,y):=\Phi(x)\Psi(y),
\end{align}
where, for instance, we choose
\begin{align}\label{s0-tanh}
\Phi(x):=\frac{s_{\max}}{2}\left[\tanh\left( \frac{x-c_1}{c_2}\right)+1\right]\chi_{[\bar{a},\bar{b}]},
\end{align}
$s_{\max}$ is the initial maximum SDF-1a concentration, $c_1$, $c_2$ are constants. Function (\ref{s0-tanh}) yields a non uniform haptotactic gradient, that is stronger at the beginning and then tends to saturate when cells acquire enough velocity. Values for $c_1$, $c_2$ will be given in Section \ref{sec:dynamic}. Then we set 
\begin{align}\label{s0-psi}
\Psi(y):=u_{\varepsilon}(y)\ast \chi_{[m-l,m+l]}(y)=\int^d_c u_{\varepsilon}(y-\tau)\chi_{[m-l,m+l]}(y)\,d\tau,
\end{align}
the convolution of $\chi_{[m-l,m+l]}(y)$ with a positive and symmetric mollifier 
\begin{align}\label{eq:mollifier}
	u_{\varepsilon}(y)&:=
	\left\{
	\begin{array}{lll}
\displaystyle\frac{1}{J}\frac{1}{\varepsilon}\;e^{-\displaystyle\frac{1}{1-\left(y/\varepsilon\right)^2}},&\text{if}&-\varepsilon<y<\varepsilon;\\\\
	0,&&\text{otherwise};
	\end{array}
	\right.
\end{align}
where 
\begin{align}\label{eq:mollifier2}
J&:=\int_{-\varepsilon}^{\varepsilon}\frac{1}{\varepsilon}\;e^{-\displaystyle\frac{1}{1-\left(y/\varepsilon\right)^2}}\,dy,
\end{align}
is the normalization factor. Mollifier (\ref{eq:mollifier}) is introduced to have sufficient smoothness for $s_0(\mathbf{x})$. A suitable value for the positive constant $\varepsilon$ will be given in Section \ref{sec:dynamic}. 
%
%adimensionalizzazione
\subsection{The nondimensional model}
Though we tend to use dimensional times and distances in the plots for easier comparison with experiments, the qualitative behaviour of the model (\ref{sys-dim}) is more clearly described using a nondimensional based on the following dimensionless quantities:
\footnotesize
\begin{equation*}
\begin{split}
  t^*:=\frac{t}{T},\quad \mathbf{x}^*:=\frac{\mathbf{x}}{R},\quad \mathbf{X}^*:=&\frac{\mathbf{X}}{R},\quad f^*:=\frac{f}{f_{\max}},\quad s^*:=\frac{s}{s_{\max}},\\
 % f^*_{\max}:=\frac{f_{\max}}{f_{\max}}=1,&\quad s^*_{\max}:=\frac{s_{\max}}{s_{\max}}=1,\\
  W^*:=\frac{W}{R^2},\quad R_i^*:=\frac{R_i}{R}\quad& i=1,\dots,5, \quad \bar{R}^*:=\frac{\bar{R}}{R},\\
\alpha^*:=\frac{\alpha s_{\max}T^2}{R^2},\quad \beta^*_\text{L}:=\beta_\text{L} T,\quad \beta^*_\text{F}:=&\beta_\text{F} T,\quad\gamma^*:=\frac{\gamma f_{\max} T^2}{R^2},\quad \omega^*_{\text{rep}}:=\frac{\omega_{\text{rep}} T^2}{R^2},\\
\omega^*_{\text{adh,L}}:=\omega_{\text{adh,L}}T^2,\quad\omega^*_{\text{adh},F}:=\omega_{\text{adh,F}}T^2,\quad\mu^*_L&:=\mu_\text{L} T,\quad \mu^*_\text{F}:=\mu_\text{F} T,\quad \delta^*:=\frac{\delta s_{\max}}{\lambda},\quad k^*_\text{L}:=\frac{k_\text{L}}{\lambda},\\
k^*_\text{F}:=\frac{k_\text{F}}{\lambda},\quad D^*:=\frac{D T}{R^2},\quad \xi^*:=\frac{\xi T}{f_{\max}},\quad \eta^*:=&\eta T,\quad \sigma^*:=\sigma T,\quad c^*_1:=\frac{c_1}{R},\quad c^*_2:=\frac{c_2}{R},\quad \varepsilon^*:=\frac{\varepsilon}{R},
\end{split}
\end{equation*}
\normalsize
where $T$ a characteristic time (see Appendix \ref{ap:parameter}). With these definitions, and maintaining for simplicity the asterisks only for the nondimensional constants, we have 
\footnotesize
\begin{equation}\label{sys-adim}
\begin{split}
&\left\{
\begin{array}{lll}
\ddot{\mathbf{X}}_i&=\displaystyle\frac{\alpha^*}{W}\int_{\mathbf{B}(\mathbf{X}_i,\bar{R}^*)}\nabla s(\mathbf{x},t)w_i(\mathbf{x})\,d\mathbf{x}+\displaystyle	\frac{\gamma^*(1-\varphi_i)}{W}\int_{\mathbf{B}(\mathbf{X}_i,\bar{R}^*)}\nabla f(\mathbf{x},t)w_i(\mathbf{x})\,d\mathbf{x}\\\\
	&+\displaystyle\frac{1}{\bar{N}_i}\sum_{j:\mathbf{X}_j\in \mathbf{B}(\mathbf{X}_i,R^*_1)\backslash\left\{\mathbf{X}_i\right\}}\mathbf{H}(\dot{\mathbf{X}}_j-\dot{\mathbf{X}}_i)+\sum_{j:\mathbf{X}_j\in \mathbf{B}(\mathbf{X}_i,R^*_5)\backslash\left\{\mathbf{X}_i\right\}}\mathbf{K}(\mathbf{X}_j-\mathbf{X}_i)-\left[\mu^*_\text{F}+(\mu^*_\text{L}-\mu^*_\text{F})\varphi_i\right]\dot{\mathbf{X}}_i,\\\\
\displaystyle\varphi_i&\displaystyle=\left\{
\begin{array}{lll}
0&\text{if}&
\displaystyle\frac{\delta^*}{W}\int_{\mathbf{B}(\mathbf{X}_i,\bar{R}^*)}s(\mathbf{x},t)w_i(\mathbf{x})\,d\mathbf{x}-\frac{k^*_\text{F}+(k^*_\text{L}-k^*_\text{F})\varphi_i}{W}\int_{\mathbf{B}(\mathbf{X}_i,\bar{R}^*)}\frac{f(\mathbf{x},t)}{1+f(\mathbf{x},t)}w_i(\mathbf{x})\,d\mathbf{x}\\\\
&&+\displaystyle\Gamma(n_i)\leq0,\\\\
1&&\text{otherwise},
\end{array}
\right.
\\\\\displaystyle	\partial_t f&\displaystyle=D^*\Delta f+\xi^*\sum_{j=1}^{N_{\text{tot}}}\displaystyle\varphi_j\chi_{\mathbf{B}(\mathbf{X}_j,R_3^*)}-\eta^* f,\\\\
\displaystyle	\partial_t s&\displaystyle=-\sigma^* s\sum_{j=1}^{N_{\text{tot}}}\displaystyle\chi_{\mathbf{B}(\mathbf{X}_j,R_3^*)},
	\end{array}
	\right.
	\end{split}
\end{equation}
\normalsize
where  
\begin{align}\label{funz-gamma-adim}
	\Gamma(n_i):=\frac{e^{n_i}}{e^{n_i}+\Gamma_0}-\frac{1}{1+\Gamma_0},
\end{align}
\begin{align}\label{n-i-adim}
	n_i:=\text{card}\left\{j:\mathbf{X}_j\in \mathring{\mathbf{B}}(\mathbf{X}_i,R^*_2)\backslash\left\{\mathbf{X}_i\right\}\right\},
\end{align}
\begin{align}
\mathbf{H}:=\left[\beta^*_\text{F}+(\beta^*_\text{L}-\beta^*_\text{F})\varphi_i\varphi_j\right]\frac{R^{*2}_1}{R^{*2}_1+||\mathbf{X}_j-\mathbf{X}_i||^2}(\dot{\mathbf{X}}_j-\dot{\mathbf{X}}_i),
\end{align}
and
\footnotesize
\begin{align}
\mathbf{K}(\mathbf{X}_j-\mathbf{X}_i):=
\left\{
\begin{array}{lll}
-\omega^*_{\text{rep}}\left(\displaystyle\frac{1}{||\mathbf{X}_j-\mathbf{X}_i||}-\frac{1}{R^*_4}\right)\displaystyle\frac{\mathbf{X}_j-\mathbf{X}_i}{||\mathbf{X}_j-\mathbf{X}_i||},&\text{if}& ||\mathbf{X}_j-\mathbf{X}_i||\leq R^*_4,\\\\
\left[\omega^*_{\text{adh,F}}+(\omega^*_{\text{adh,L}}-\omega^*_{\text{adh,F}})\varphi_i\varphi_j\right]\left(||\mathbf{X}_j-\mathbf{X}_i||-R^*_4\right)\displaystyle\frac{\mathbf{X}_j-\mathbf{X}_i}{||\mathbf{X}_j-\mathbf{X}_i||},&\text{if}& R^*_4<||\mathbf{X}_j-\mathbf{X}_i||\leq R^*_5.
\end{array}
\right.
\end{align}%
\normalsize
Initial and boundary conditions are still given by (\ref{pos-vel-iniz})--(\ref{s0-psi}). In particular in (\ref{s0-tanh}), (\ref{eq:mollifier}), and (\ref{eq:mollifier2}) we have to replace $s_{\max}$, $c_1$, $c_2$ and $\varepsilon$, with $s^*_{\max}=1$, $c^*_1$, $c^*_2$ and $\varepsilon^*$.
\section{Steady states and stability}\label{sec:steady}
\begin{figure}[!ht]
\centering
\subfigure[]
{
\begin{pspicture}(-3,-3)(3,3)
\uput[90](0,-0.3){$\mathbf{X}_\text{L}$}
\pscircle(0,0){1}
\pscircle(1.5,0){1}
\pscircle(-1.5,0){1}
\pscircle(0,1.5){1}
\uput[90](0.05,-1.95){$\mathbf{X}_i$}
\pscircle(0,-1.5){1}
\pscircle(1.05,1.05){1}
\pscircle(-1.05,-1.05){1}
\pscircle(1.05,-1.05){1}
\pscircle(-1.05,1.05){1}
\end{pspicture}
}
\hspace{0 mm}
\subfigure[]
{
\begin{pspicture}(-3,-4)(3,1)
\psline{->}(0,0)(1.5,0)
\uput[270](1.3,0){$x$}
\psline{->}(0,0)(0,1.5)
\uput[180](0,1.3){$y$}
\psarc(0,0){.4}{225}{270}
\psline(0,0)(0,-3)
\psarc(0,-3){.3}{90}{157.5}
\psline(0,0)(2.121,-2.121)
\psline(0,0)(-2.121,-2.121)
\psarc(-2.121,-2.121){.6}{337.5}{360}
\psline(0,-3)(2.121,-2.121)
\psline(0,-3)(-2.121,-2.121)
\psline[linestyle=dashed](-2.121,-2.121)(2.121,-2.121)
\uput[248](-0.1,-0.3){$\alpha_1$}
\uput[351](-1.5,-2.3){$\alpha_3$}
\uput[100](-0.2,-2.8){$\alpha_2$}
\uput[145](0.1,0){$\mathbf{X}_\text{L}$}
\uput[180](-2.121,-2.121){$\mathbf{X}_{i-1}$}
\uput[360](2.121,-2.121){$\mathbf{X}_{i+1}$}
\uput[270](0,-3){$\mathbf{X}_i$}
\uput[135](-0.9,-1.06){$d_1$}
\uput[180](0.1,-1.5){$d_1$}
\uput[270](-1.06,-2.5){$d_3$}
\uput[90](0.5,-2.18){$d_2$}
\end{pspicture}
}
\caption{(a) Example of 8-rosette with a leader centered in $\mathbf{X}_\text{L}$ and 8 followers centered in $\mathbf{X}_i$, $i=1,\dots,8$. (b) Geometrical configuration of a $N$-rosette with a leader cell centered in $\mathbf{X}_\text{L}$ and some followers centered in $\mathbf{X}_{i-1}$, $\mathbf{X}_{i}$, $\mathbf{X}_{i+1}$.}
\label{fig-8rosette}
\end{figure}
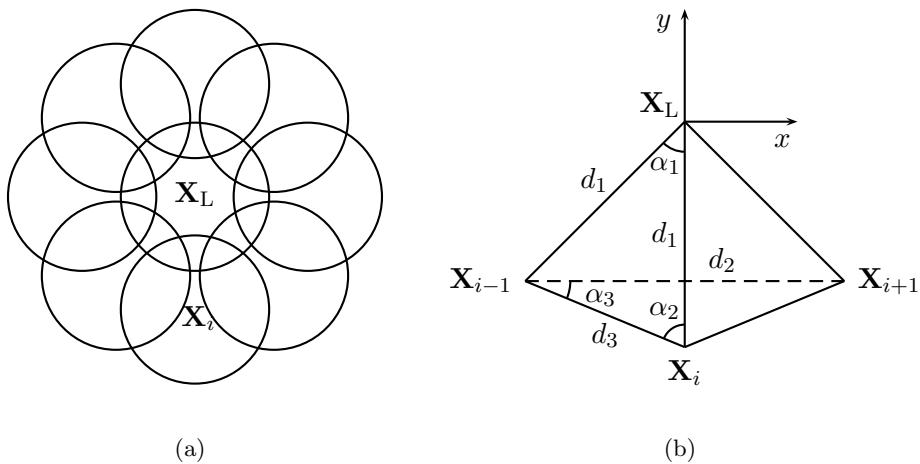

Now we will investigate particular steady states for our model. They are biologically relevant, because they correspond to the neuromasts basic structure (see Section \ref{sec:bio}). This will be useful also to provide us with a range of variability for some parameters or to specify some of their ratios. First we consider the stationary form of system (\ref{sys-adim})
\begin{equation}\label{sys-adim-staz}
\begin{split}
&\left\{
\begin{array}{l}
\displaystyle\frac{\gamma^*(1-\varphi_i)}{W}\int_{\mathbf{B}(\mathbf{X}_i,\bar{R}^*)}\nabla f(\mathbf{x})w_i(\mathbf{x})\,d\mathbf{x}+\sum_{j:\mathbf{X}_j\in \mathbf{B}(\mathbf{X}_i,R^*_5)\backslash\left\{\mathbf{X}_i\right\}}\mathbf{K}(\mathbf{X}_j-\mathbf{X}_i)=\mathbf{0},\\\\
\displaystyle\varphi_i=\left\{
\begin{array}{lll}
0&\text{if}&
-\displaystyle\frac{k^*_\text{F}+(k^*_\text{L}-k^*_\text{F})\varphi_i}{W}\int_{\mathbf{B}(\mathbf{X}_i,\bar{R}^*)}\frac{f(\mathbf{x})}{1+f(\mathbf{x})}w_i(\mathbf{x})\,d\mathbf{x}+\Gamma(n_i)\leq0,\\\\
1&&\text{otherwise},
\end{array}
\right.\\\\
D^*\Delta f=\eta^* f-\xi^*\displaystyle\sum_{j=1}^{N_{\text{tot}}}\displaystyle\varphi_j\chi_{\mathbf{B}(\mathbf{X}_j,R_3^*)},\\\\
s=0,
	\end{array}
	\right.
	\end{split}
\end{equation}
with
\begin{align*}
\frac{\partial f}{\partial\mathbf{n}}=0,\quad\mbox{on $\partial\Omega$}.
\end{align*}
\begin{Definition}
We will call \emph{$N$-rosette} ($N\geq 2$) a configuration formed by a leader cell surrounded by $N$ follower cells with their centers located on the vertices of a regular polygon of $N$ sides (or a segment if $N=2$) centered in the leader cell (Figure \ref{fig-8rosette} (a)).
\end{Definition}
\par With reference to Figure \ref{fig-8rosette} (b), we call $\mathbf{X}_\text{L}$ the center of the leader cell, $\mathbf{X}_i$ the center of a follower, $d_1$ the distance between the followers and the leader, $d_2$ the distance between two followers in alternating position (e.g. $\mathbf{X}_{i-1}$ and $\mathbf{X}_{i+1}$), $d_3$ the distance between two adjoining followers (e.g. $\mathbf{X}_{i}$ and $\mathbf{X}_{i+1}$), and $\alpha_1$, $\alpha_2$, $\alpha_3$ the angles in the figure.  
\par By symmetry considerations we set:
\begin{align}
\alpha_1=\frac{2\pi}{N},\quad\alpha_2&=\frac{\pi-\alpha_1}{2},\quad\alpha_3=\frac{\pi}{2}-\alpha_2=\frac{\pi}{N},\\
d_2&=2d_1\sin \frac{2\pi}{N},\label{dist-d2}\\
d_3&=2d_1\sin \frac{\pi}{N}.\label{dist-d3}
\end{align}
Now we assume the following physically reasonable hypothesis for a $N$-rosette:
\begin{itemize}
 \item the range of lateral inhibition equal to the range of repulsion between cells:
\begin{align}\label{hp1}
	R_2=R_4;
\end{align}
\item the followers are located in the range of the lateral inhibition of the leader:
\begin{align}\label{hp2}
	d_1\leq R_2;
\end{align}
\item there is no repulsion between adjoining followers if $2\leq N\leq 4$:
\begin{align}\label{hp3A}
	d_3\geq R_4;
\end{align}
there is no repulsion between followers in alternating position if $N\geq 5$:
\begin{align}\label{hp3B}
d_2\geq R_4.
\end{align}
\end{itemize}
We point out that hypothesis \eqref{hp2} is a direct consequence of the definition of a $N$-rosette. 
\par Taking system (\ref{sys-adim-staz}) and hypothesis \eqref{hp1}--\eqref{hp3B} into account, we can state the following results.
\begin{Proposition}\label{prop-nrosette}
There exist $N$-rosettes if and only if $N\leq 12$. Moreover the distance $d_1$, depending on $N$, can vary in the following ranges:
\begin{align}
	\frac{1}{2\sin \frac{\pi}{N}}\leq &\frac{d_1}{R_4}\leq 1,\quad\text{if}\quad 2\leq N\leq 4,\label{prop-distA}\\
	\frac{1}{2\sin \frac{2\pi}{N}}\leq &\frac{d_1}{R_4}\leq 1,\quad\text{if}\quad 5\leq N\leq 12.\label{prop-distB}
\end{align}
\end{Proposition}
\emph{Proof.}
Condition (\ref{prop-distA}) is a consequence of (\ref{hp2}), (\ref{hp1}), (\ref{hp3A}), and (\ref{dist-d3}). While (\ref{prop-distB}) is a consequence of (\ref{hp2}), (\ref{hp1}), (\ref{hp3B}), and (\ref{dist-d2}). In particular (\ref{prop-distB}) is not empty if and only if $N\leq 12$. 
\bigskip
\par The maximum number of cells, which is provided by the previous proposition, is consistent with the experimental observations as shown in \cite{gilmour}. 
%
%prop 2
\begin{Proposition}\label{prop-repulsion}
In a $N$-rosette there are repulsion and lateral inhibition effects between adjoining followers if and only if $N\geq 5$. In particular if $N\geq 6$ these effects do not depend on $d_1$, and if $N=5$ this holds if and only if
\begin{align}\label{rep-n5}
	\frac{1}{2\sin \frac{2\pi}{5}}\leq &\frac{d_1}{R_4}<\frac{1}{2\sin \frac{\pi}{5}}.
\end{align}
\end{Proposition}
\emph{Proof.}
Hypothesis (1) ensures that this proof holds both for repulsion and lateral inhibition effects. If $N\leq 4$ the statement is  true thanks to hypothesis (2).
\\ If $N\geq 6$ from (\ref{dist-d3}), (\ref{hp1}), and (\ref{hp2}) we have 
\begin{align*}
	d_3\leq d_1\leq R_2=R_4,
\end{align*}
independently from $d_1$.
\\If $N=5$, using (\ref{dist-d3}) we have repulsion if and only if 
\begin{align}\label{rep-n5-cond}
	d_3=2 d_1\sin\frac{\pi}{5}<R_4.
\end{align}
From (\ref{rep-n5-cond}) and (\ref{prop-distB}) equation (\ref{rep-n5}) follows .
\bigskip
\par Now, in order to fix the range of variability for some parameters we solve the nondimensional system (\ref{sys-adim-staz}) for a $N$-rosette with a leader in $\mathbf{X}_\text{L}$, setting a frame centered $\mathbf{X}_\text{L}$ with axes passing through the center of a follower (Figure \ref{fig-8rosette} (b)). For simplicity we introduce the following symbols:
\begin{align*}
	\overline{\nabla f(\mathbf{X}_i)}&:=\frac{1}{W}\int_{\mathbf{B}(\bar{R}^*,\mathbf{X}_i)}\nabla f(\mathbf{x})w_i(\mathbf{x})\,d\mathbf{x},\\
\overline{\overline{f(\mathbf{X}_i)}}&:=\frac{1}{W}\int_{\mathbf{B}(\bar{R}^*,\mathbf{X}_i)}\frac{ f(\mathbf{x})}{1+f(\mathbf{x})}w_i(\mathbf{x})\,d\mathbf{x},
\end{align*} 
to denote the weighted average of the functions $\nabla f$ and $\frac{f}{1+f}$.
\par Firstly, equation (\ref{sys-adim-staz})$_2$ for each followers and for the leader becomes respectively:
\begin{align}
\varphi_i&=0 \Leftrightarrow -k^*_\text{F} \overline{\overline{f(\mathbf{X}_i)}} +\Gamma(n)\leq 0,\quad i=1,\dots,N, \label{phi-i-staz}\\
\varphi_0&=1 \Leftrightarrow -k^*_\text{L} \overline{\overline{f(\mathbf{X}_\text{L})}} +\Gamma(N)>0,\label{phi-0-staz}
\end{align}
where the function $\Gamma$ is given by (\ref{funz-gamma-adim}). Here the number $n$, which is related to the lateral inhibition, is given by (\ref{n-i-adim}) and, by symmetry considerations, it does not depend on $i$. Moreover, according to hypothesis (2) and Proposition \ref{prop-repulsion}, it takes only the values 1 or 3: if $N\geq 6$, or $N=5$ and holds condition (\ref{rep-n5}), we have to take $n=3$, otherwise $n=1$. The case $n=3$ means that on the $i$-th cell we have the lateral inhibition of the leader cell and of the two adjoining followers, while in the case $n=1$ we have only the lateral inhibition of the leader cell. The other cases for $n$ are not possible due to conditions \eqref{hp1}--\eqref{hp3B} assumed on the distances.
\par Now, the function $f(\mathbf{x})$, which is needed in (\ref{phi-i-staz}) and (\ref{phi-0-staz}), is the solution in the domain $\Omega$ of equation (\ref{sys-adim-staz})$_3$, with Neumann boundary conditions, that in this case it takes the form
\begin{align}
	D^*\Delta f-\eta^* f&=-\xi^*\chi_{\mathbf{B}(\mathbf{X}_\text{L},R_3^*)},\label{eq:f:adim:staz}\\
	\frac{\partial f}{\partial\mathbf{n}}&=0,\quad \text{on $\partial\Omega$},\label{eq:f:adim:staz2}
\end{align}
$\mathbf{X}_\text{L}$ being the center of the leader cell, the only one that produces FGF signal. If $\Omega$ is a circular domain centered in the leader cell, radial symmetry of the solution of (\ref{eq:f:adim:staz}) and (\ref{eq:f:adim:staz2}) implies the quantities $\overline{\overline{f(\mathbf{X}_i)}}$ to be the same for all $i$, so that (\ref{phi-i-staz}) and (\ref{phi-0-staz}) become
\begin{align}
 k^*_\text{F}&\geq  \bar{k}^*_\text{F}:=\Gamma(n)/\overline{\overline{f(\mathbf{X}_i)}},\label{kf-cond}\\
 k^*_\text{L}&<\bar{k}^*_\text{L}:=\Gamma(N)/\overline{\overline{f(\mathbf{X}_\text{L})}}.\label{kl-cond}
\end{align}
\par Now we try to obtain a numerical estimate for the bound functions $\bar{k}^*_\text{F}$ and $\bar{k}^*_\text{L}$, as $N$ changes. We set a domain $\Omega=[0,200]\times[0,200]\;(\mu\text{m}^2)$ with a single leader cell located in $\mathbf{X}_\text{L}=(100,100)\;(\mu\text{m})$. We choose the square domain size sufficiently large, so that its influence on the solution can be neglected in the time period of interest. Then equations (\ref{eq:f:adim:staz}) and (\ref{eq:f:adim:staz2}) are numerically solved in such a domain, as described in Section \ref{ap:scheme} to follow, using a spatial discretization corresponding to $\Delta x=\Delta y=0.2\;\mu\text{m}$. Parameters $D^*$, $\eta^*$, $\xi^*$, $R^*_2$, $R^*_3$, $R^*_4$, $\bar{R}^*$ used here are listed in Table \ref{tab-par-adim}, Appendix \ref{ap:parameter}.  
\par
\begin{figure}[!ht]
\centering
\subfigure[]
{\includegraphics[width=0.5\textwidth]{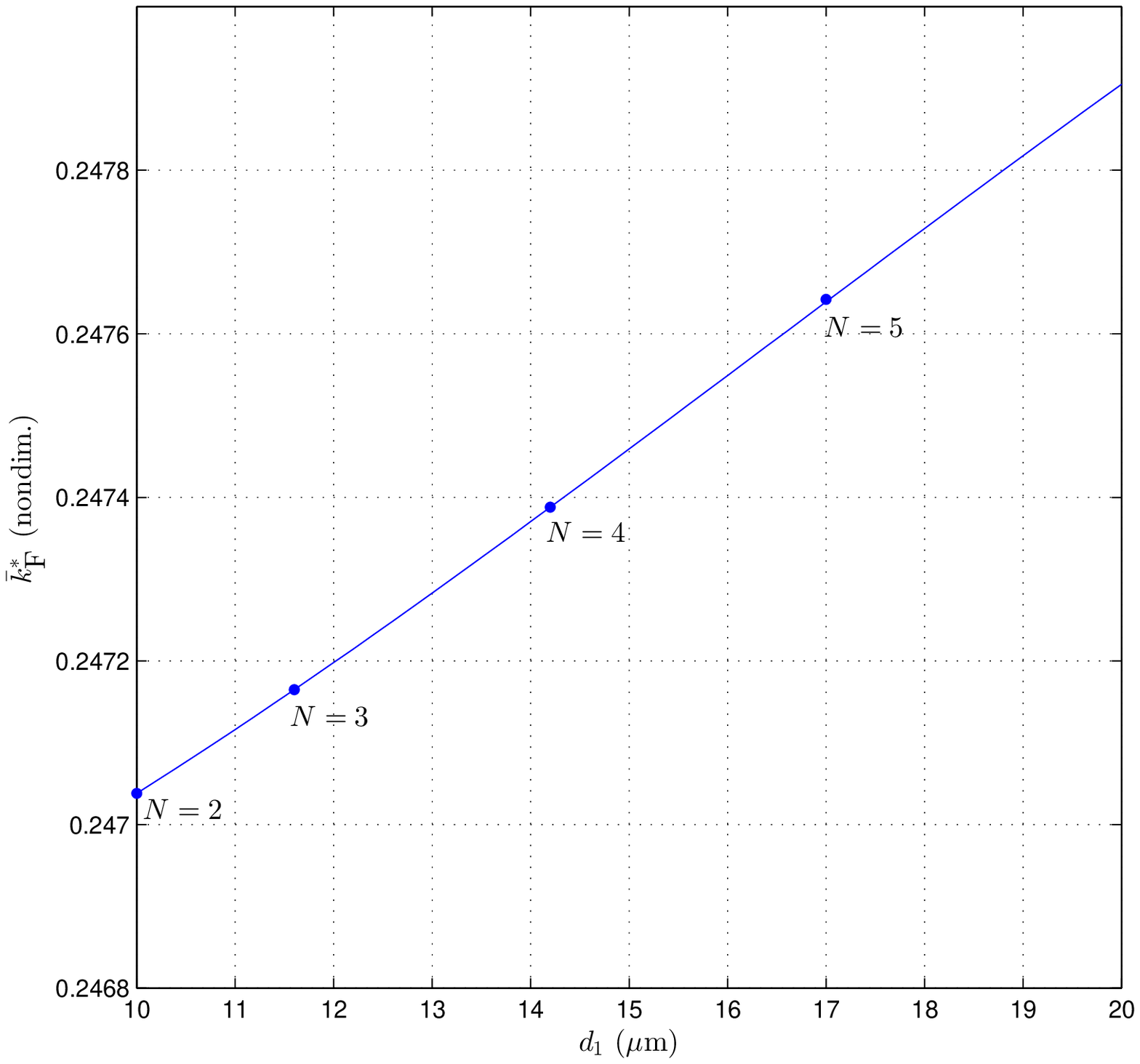}}
\hspace{-9 mm}
\subfigure[]
{\includegraphics[width=0.5\textwidth]{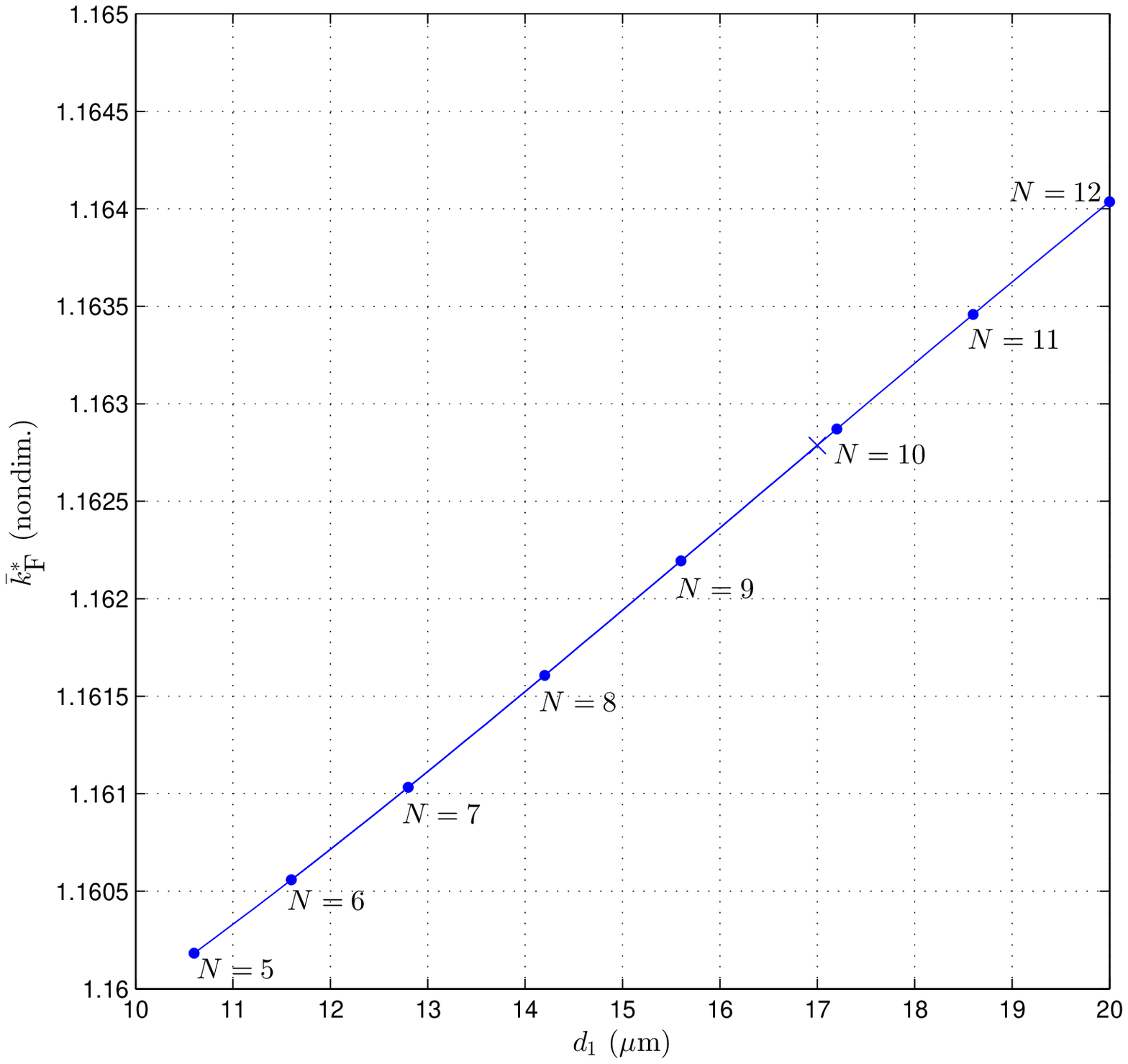}}
\caption{Numerical plot of the lower bound $\bar{k}^*_\text{F}$  in (\ref{kf-cond}) as a function of $d_1$. The curve gives the lower bound for ${k}^*_\text{F}$ for a fixed $N$ and $d_1$. Since the scale of the curves is essentially different, we present in (a) the case $N=2,3,4$, in which the curves are overlapped, and similarly in (b) the case $N=6,\dots,12$. The starting point on the curves is marked by ``$\bullet$'', while the ending point is represented by $d_1=R_4$ for all $N$. For $N=5$ the curve starts in (b) until $d_1\approx 17\;\mu\text{m}$ (marker ``$\times$''), then for larger values of $d_1$ continues in (a). For $N=12$ the curve is reduced to a single value in $d_1=R_4$ in (b).}
\label{fig:K}
\end{figure}

 Figure \ref{fig:K} shows a numerical estimate for the lower bound $\bar{k}^*_\text{F}$ in (\ref{kf-cond}). For each fixed value of $N$, $N=2,\dots,12$, the curve indicates the value of $\bar{k}^*_\text{F}$ as a functions of  $d_1$, which is the distance between leader and follower. Since the scale of the curves is essentially different as $N$ changes, we present our results in two different pictures in Figures \ref{fig:K}. Notice that the range of the distance $d_1$ to be considered depends on $N$ according to (\ref{prop-distA}) and (\ref{prop-distB}) in Proposition \ref{prop-nrosette}: the starting point on the curve is marked by a ``$\bullet$'', while the ending point is represented by $d_1=R_4$ for all $N$. We have already observed that $\Gamma(n)$ can only obtain the values $\Gamma(1)$ or $\Gamma(3)$ according to Proposition \ref{prop-repulsion}. So, clearly, the curves for $N=2,3,4$ are overlapped (Figure \ref{fig:K} (a)), the same for $N=6,\dots,12$ (Figure \ref{fig:K} (b)). For $N=5$ the curve starts in Figure \ref{fig:K} (b), with $\Gamma(n)=\Gamma(3)$, until $d_1\approx 17\;\mu\text{m}$ (marker ``$\times$''), then $\Gamma(n)$ becomes $\Gamma(1)$ and, for larger values of $d_1$, the curve continues in Figure \ref{fig:K} (a). For $N=12$ the right hand side of (\ref{kf-cond}) assumes a single value in $d_1=R_4$ in Figure \ref{fig:K} (b).
\par On the other hand, using again the numerical solution of $f(\mathbf{x})$, the right hand side of (\ref{kl-cond}) can be tabulated as $N$ changes. Its values are given in Table \ref{tab:K}. 
\begin{table}[!ht]
\caption{Numerical values of the upper bound $\bar{k}^*_\text{L}$ in (\ref{kl-cond}) for $N=2,\dots,12$. In practice fixing $N$ we have the upper bound of $k^*_\text{L}$ for the existence of a steady $N$-rosette.}
\label{tab:K}
\centering
\begin{tabular}{||c|c||c|c||}
\hline
\rule[-3 mm]{0 mm}{0.9 cm}
$N$&$ \bar{k}^*_\text{L}$ (nondim.)&$N$&$ \bar{k}^*_\text{L}$ (nondim.)\\
\hline
\hline
$2$&0.6707&8&1.8187\\
\hline 
$3$&1.1580&9&1.8230\\
\hline
$4$&1.5146&10&1.8245\\
\hline
$5$&1.6987&11&1.8251\\
\hline 
$6$&1.7769&12&1.8253\\
\cline{1-2}
$7$&1.8073&&\\
\hline 
\hline 
\end{tabular}
\end{table}
\par
\begin{figure}[!ht]
	\centering
		\includegraphics[width=0.7\textwidth]{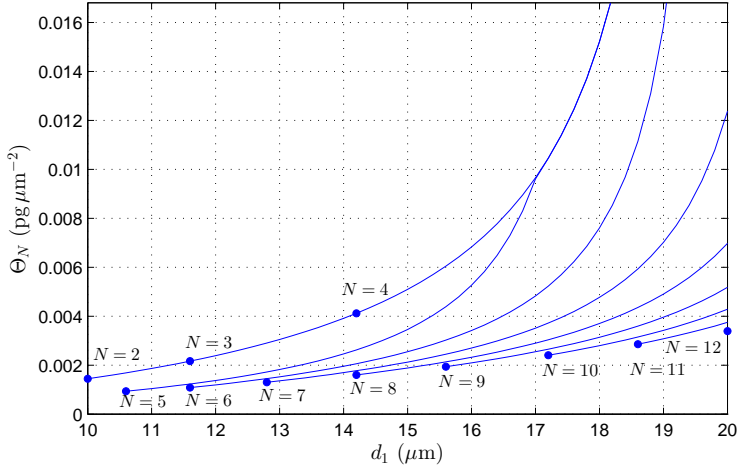}
	\caption{Dimensional numerical plot of $\Theta_N$ for $N=2,\dots,12$, that demonstrates that this function is monotone with respect to $d_1$ and then invertible. The first three curves, for $N=2,\dots,4$, coincide (first line in the top). Then, from the top to the bottom, we have the curves related to $N=5,\dots,11$. For $N=5$ the curve goes to coincide with the first curve on the top about from $d_1>17\;\mu\text{m}$. For $N=12$ the curve is reduced to a single value in $d_1=R_4$. In practice fixing $N$ and $d_1$ we have the value of $\Theta_N$ that provides in (\ref{omega-gamma-d}) the ratio of the parameters $\omega^*_{\text{rep}}$ and $\gamma^*$.}
		\label{fig:curve-monotonia}
\end{figure}

 Now, equation (\ref{sys-adim-staz})$_1$ becomes
\begin{align}
&\sum_{j:\mathbf{X}_j\in \mathbf{B}(\mathbf{X}_\text{L},R^*_5)\backslash\left\{\mathbf{X}_\text{L}\right\}}\mathbf{K}(\mathbf{X}_j-\mathbf{X}_\text{L})=\mathbf{0},\label{Ieq-staz-leader}\\
&\gamma^*\overline{\nabla f(\mathbf{X}_i)}+\sum_{j:\mathbf{X}_j\in \mathbf{B}(\mathbf{X}_i,R^*_4)\backslash\left\{\mathbf{X}_i\right\}}\mathbf{K}(\mathbf{X}_j-\mathbf{X}_i)=\mathbf{0},\quad i=1,\dots,N,\label{Ieq-staz-foll}
\end{align}
respectively for the leader and for each follower. Here $\mathbf{K}$ contains only repulsion term:
\begin{align*}
\mathbf{K}(\mathbf{X}_j-\mathbf{X}_i):=-\omega^*_{\text{rep}}\left(\frac{1}{||\mathbf{X}_j-\mathbf{X}_i||}-\frac{1}{R^*_4}\right)\frac{\mathbf{X}_j-\mathbf{X}_i}{||\mathbf{X}_j-\mathbf{X}_i||}.
\end{align*}
For symmetry (\ref{Ieq-staz-leader}) is identically satisfied. Then in (\ref{Ieq-staz-foll}) $f(\mathbf{x},t)$ is given by (\ref{eq:f:adim:staz}) and (\ref{eq:f:adim:staz2}), so in a circular domain $\Omega$ we can write the same relation for all $i$. For example, in relation to Figure \ref{fig-8rosette} (b), we have 
\begin{align}\label{dist-equil}
\gamma^*\overline{\partial_y f(\mathbf{X}_i)}-\omega^*_{\text{rep}}h_1(d^*_1)-\omega^*_{\text{rep}}h_2(N,d^*_1)=0,
\end{align}    
in which $\omega^*_{\text{rep}}h_1(d^*_1)$ represents the repulsion of the leader:
\begin{align*}
	h_1(d^*_1):=\frac{1}{d^*_1}-\frac{1}{R^*_4},
\end{align*}
$d^*_1=d_1/R$ is the nondimensional value of $d_1$, and $\omega^*_{\text{rep}}h_2(N,d^*_1)$ is the possible repulsion of two adjoining followers according to (\ref{dist-d3}) and Propositions \ref{prop-nrosette}, \ref{prop-repulsion}, namely:
\begin{align}\label{h-dist-d1}
	h_2(N,d^*_1):=
\left\{
\begin{array}{lll}
	0,&\text{if}& N\leq4;\\
2\displaystyle\left(\frac{1}{2d^*_1\sin\frac{\pi}{5}}-\frac{1}{R^*_4}\right)\displaystyle\sin \frac{\pi}{5},&\text{if}& N=5 \wedge \frac{d^*_1}{R^*_4}< \frac{1}{2\sin \frac{\pi}{5}};\\
	0,&\text{if}& N=5 \wedge \frac{1}{2\sin \frac{\pi}{5}}\leq \frac{d^*_1}{R^*_4}\leq 1;\\
2\displaystyle\left(\frac{1}{2d^*_1\sin\frac{\pi}{N}}-\frac{1}{R^*_4}\right)\displaystyle\sin \frac{\pi}{N},
	&\text{if}& 6\leq N\leq 12.
\end{array}
\right.
\end{align}
We remark that equation (\ref{dist-equil}) is useful for two reasons. First, if we know an experimental value for the distance $d^*_1$ we can obtain, fixing $N$, the ratio $\omega^*_{\text{rep}}/\gamma^*$ as a function of $d^*_1$:
\begin{align}\label{omega-gamma-d}
	\frac{\omega^*_{\text{rep}}}{\gamma^*}=\Theta_N(d^*_1):=\frac{\overline{\partial_y f(\mathbf{X}_i)}}{h_1(d^*_1)+h_2(N,d^*_1)}.
\end{align}
On the other hand, if $\Theta_N$ is invertible, we can express $d^*_1$ as a function of $\omega^*_{\text{rep}}$ and $\gamma^*$ that is the equilibrium distance for a $N$-rosette fixed the physical parameters. Figure \ref{fig:curve-monotonia} represents a dimensional numerical plot of $\Theta_N$ for $N=2,\dots,12$. It shows that $\Theta_N$ is monotone with respect to $d_1$ for all $N$, so that relation (\ref{omega-gamma-d}) is invertible. To obtain this plot the value of $f(\mathbf{x})$ has been obtained numerically from (\ref{eq:f:adim:staz}) and (\ref{eq:f:adim:staz2}) as previously described, fixing the same domain and the same parameters. 
\par The domain of the curves, as in Figure \ref{fig:K}, is given by (\ref{prop-distA}) and (\ref{prop-distB}); now it represents the admissible distances $d_1$ for a $N$-rosette, as $N$ changes. Symbol ``$\bullet$'' marks the origin of the curves. For $N=2,3,4$ the curves are overlapped (first line in the top) because for them $h_2=0$ (see (\ref{omega-gamma-d}) and (\ref{h-dist-d1})). For $N=5$ the curve coincide with the curve $N=2,3,4$ when $h_2$ becomes zero. This happens about for $d_1>17\;\mu\text{m}$, as we can see in (\ref{h-dist-d1})$_{1,2,3}$. The curves corresponding to $N=2,\dots,6$ have a vertical asymptote in $d_1=R_4$ where the functions $h_1$ and $h_2$ in (\ref{omega-gamma-d}) become zero. Conversely, for $N=7,\dots,12$, $\Theta_N$ is defined in $d_1=R_4$. In particular for $N=12$, due to (\ref{prop-distB}), the curve is reduced to a single value in $d_1=R_4$ given by  $\Theta_{12}(R_4)$ (marker point on the right).
%
%
%FIGURA
\par Typical values for $N$ and $d_1$ (or $d^*_1$) to be used in (\ref{kf-cond}), (\ref{kl-cond}), and (\ref{omega-gamma-d}) will be given in Appendix \ref{ap:parameter}.
%
%
%STAZIONARIO
%
%
\par Now, in order to test numerically a steady $N$-rosette, we perform a dynamic simulation of the model (\ref{sys-adim}), as described in Section \ref{ap:scheme}, with initial data given by a solution of the stationary system (\ref{sys-adim-staz}). In particular, we consider the spatial domain $\Omega=[0,200]\times[0,200]\;(\mu\text{m}^2)$ and the time interval $[0,50]\;(\text{h})$, that is a typical time range used in the experimental observations \cite{nechiporuk}. Spatial and temporal discretizations are respectively $\Delta x=\Delta y=0.2\,\mu\text{m}$ and $\Delta t=0.01\,\text{h}$. Initial data are set as follows:
\begin{align}\label{X0staz}
	\mathbf{X}_i(0)=\mathbf{X}_{i0},
\end{align}
$\mathbf{X}_{i0}$ being a $8$-rosette centered in $\mathbf{X}_\text{L}(0)=(100,100)\;(\mu\text{m})$, with follower-leader distance fixed at $d_1=\frac{3}{2}R$ (see Appendix \ref{ap:parameter} and Figure \ref{fig:staz} (a)), 
\begin{align}\label{Xdot0staz}
	\dot{\mathbf{X}}_i(0)=\mathbf{0}, 
\end{align}
\begin{align}\label{f0staz}
	f(\mathbf{x},0)=f_0(\mathbf{x}),
\end{align}
with $f_0(\mathbf{x})$ solution of equation (\ref{sys-adim-staz})$_3$ with homogeneous Neumann boundary condition in the same domain;
\begin{align}\label{s0staz}
	s(\mathbf{x},0)=0,
\end{align}
according to (\ref{sys-adim-staz})$_4$. The parameters used here are listed in Appendix \ref{ap:parameter} (see Tables \ref{tab-param-dim}, \ref{tab-par-adim}).
\par We see that our numerical results demonstrate that, with good approximation, the initial configuration stays constant in time. Figure \ref{fig:staz} shows evolution in space of the dimensional solution at two different time steps: $t=0\;\text{h}$ and $t=50\;\text{h}$. Green colour marks the leader cell ($\varphi_i=1$), and red colour marks a follower cell ($\varphi_i=0$). Contour plot in the background is related to the FGF signal concentration, while variable $s(\mathbf{x},t)$ is not shown. Figure \ref{fig:staz_err} shows the evolution in time of the maximum relative error on the position, $E_{\text{max,rel}}(t):=\frac{\max_{1\leq i\leq N_{\text{tot}}}\left\|\mathbf{X}_i(t)-\mathbf{X}_{i0}\right\|}{R}$, and the maximum velocity $V_{\text{max}}(t):=\max_{1\leq i\leq N_{\text{tot}}}\left\|\dot{\mathbf{X}}_i(t)\right\|$. $E_{\text{max,rel}}$ suggests a deviation from the initial position in the order of $10^{-3}$ times cell radius, while $V_{\text{max}}$ is in the order of $10^{-4}$ $\mu\text{m}\,\text{h}^{-1}$, which is very small with respect to the cell velocity during migration that is around $69\;\mu\text{m}/\text{h}$ \cite{gilmour}.   
\begin{figure}[!ht]
\centering
\subfigure[]{
\includegraphics[width=0.4\textwidth]{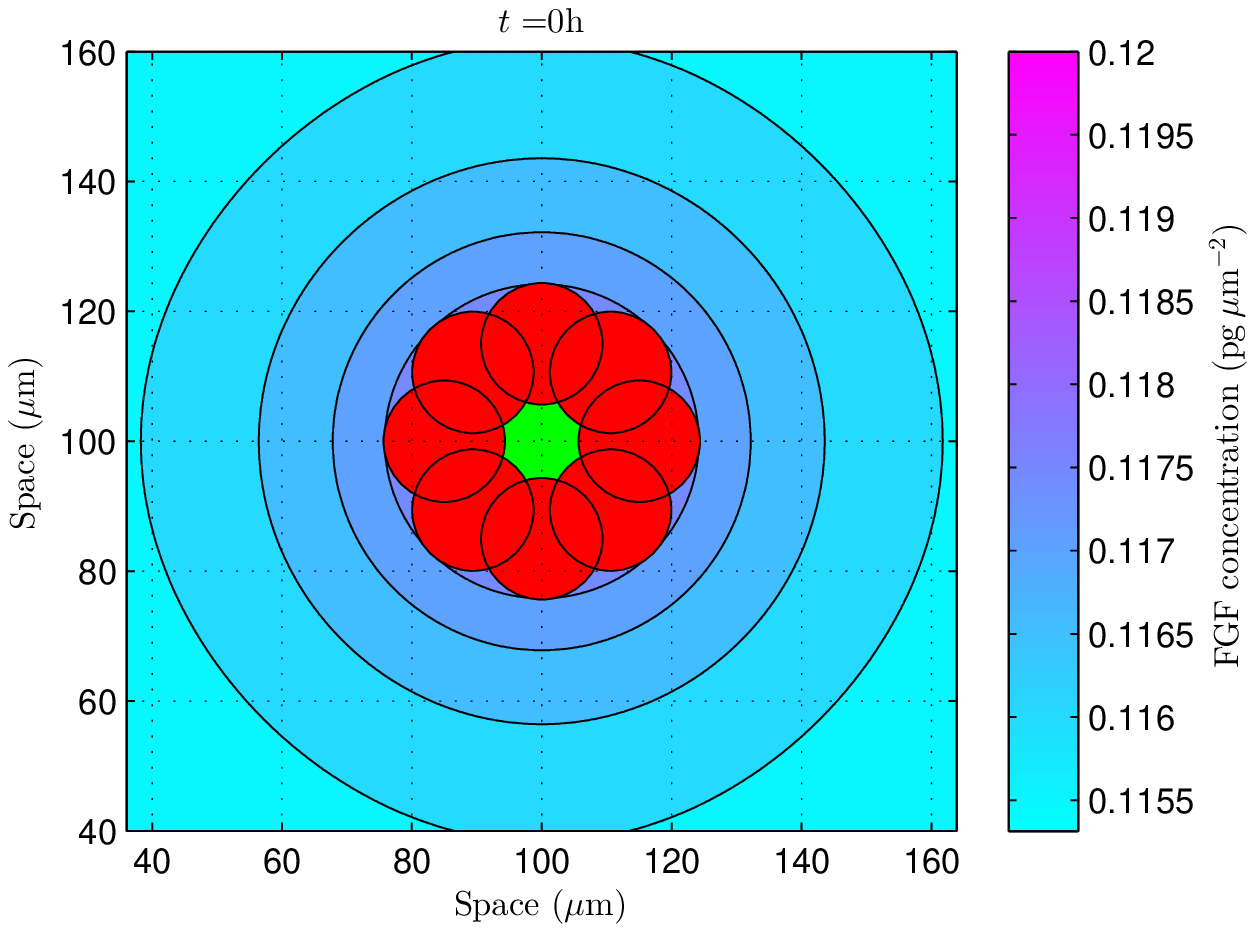}}
\hspace{0 mm}
%\subfigure[]{
%\includegraphics[width=0.4\textwidth]{staz_t16.eps}}\\
%\vspace{0 mm}
%\subfigure[]{
%\includegraphics[width=0.4\textwidth]{staz_t32.eps}}
%\hspace{0 mm}
\subfigure[]{
\includegraphics[width=0.4\textwidth]{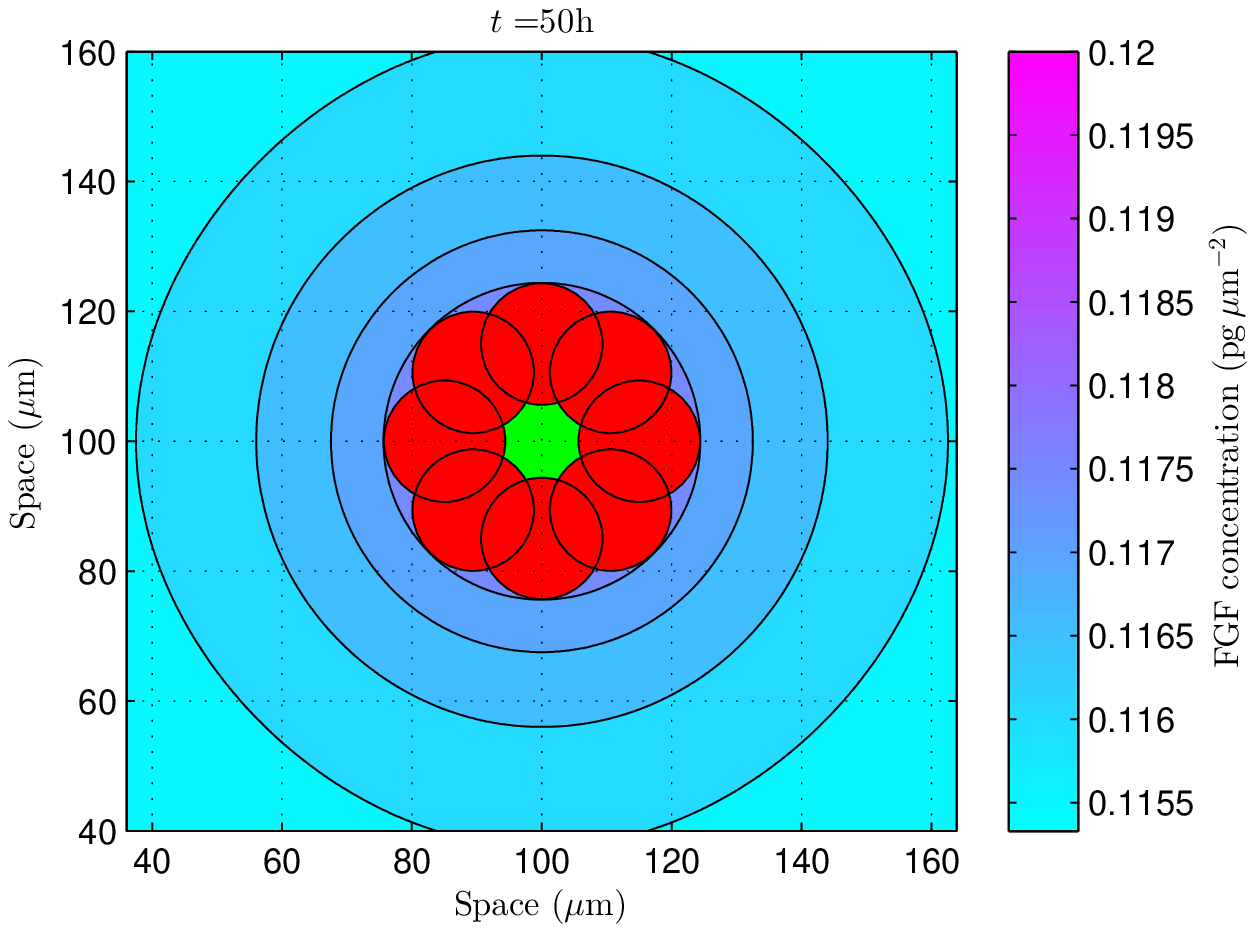}}
\caption{Numerical simulation of a steady solution given by a $8$-rosette. (a)-(b) are related respectively to the dimensional plot at two different time steps: $t=0\;\text{h}$ and $t=50\;\text{h}$. System (\ref{sys-adim}) is solved as described in Section \ref{ap:scheme} in $\Omega=[0,200]\times[0,200]\;(\mu\text{m}^2)$ (plot refers only to a part of the domain) and in $[0,50]\;(\text{h})$. Spatial and temporal discretization is set to $\Delta x=\Delta y=0.2\;\mu\text{m}$ and $\Delta t=0.01\;\text{h}$. Initial data are given by (\ref{X0staz})--(\ref{s0staz}). The parameters used here are listed in Appendix \ref{ap:parameter}. Green colour ({\large{\color{green}$\bullet$}}) marks the leader cell, red colour ({\large{\color{red}$\bullet$}}) a follower cell, contour plot in the background is the FGF signal concentration. Variable $s(\mathbf{x},t)$ is not shown.}
\label{fig:staz}
\end{figure}
\begin{figure}[!ht]
\centering
\subfigure[]{\includegraphics[width=0.48\textwidth]{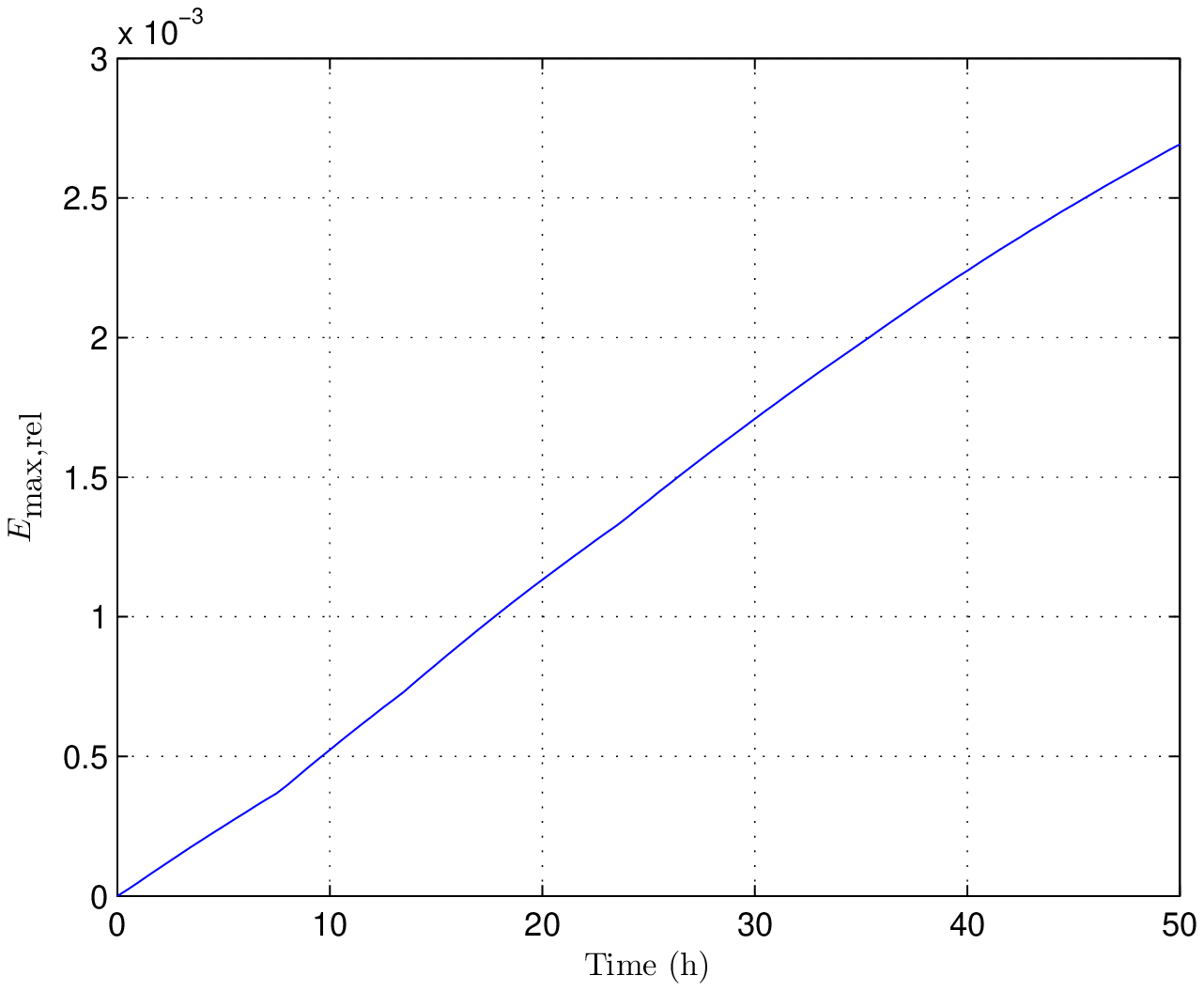}}
\hspace{1 mm}
\subfigure[]{\includegraphics[width=0.48\textwidth]{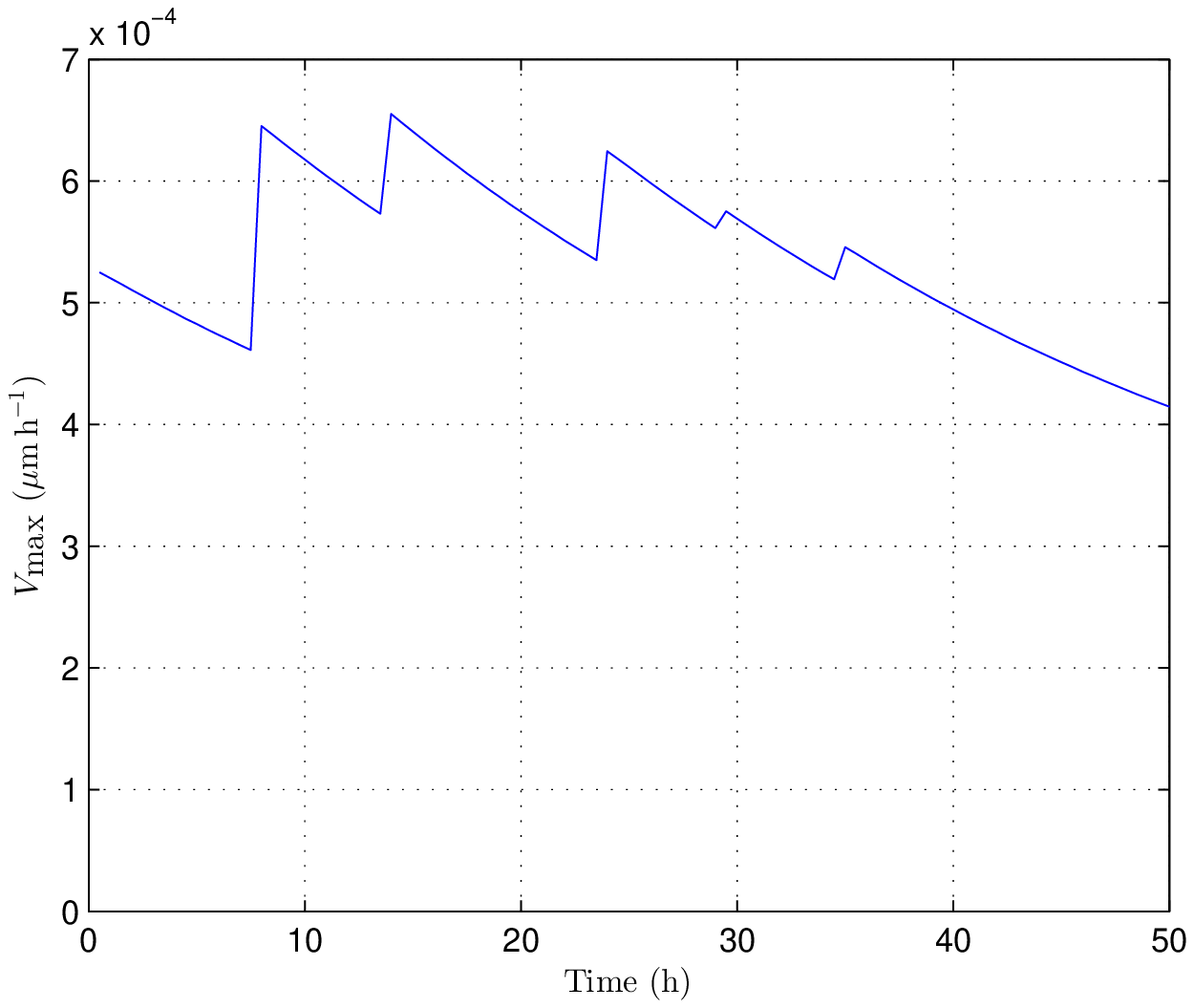}}
\hspace{1 mm}
\caption{Numerical assessment of a steady $8$-rosette.(a) Plot in time of the maximum relative error $E_{\text{max,rel}}(t)$. (b) Plot of the maximum velocity $V_{\text{max}}(t)$.}
\label{fig:staz_err}
\end{figure}
%
%
%
%
%STABILITA':
%
%
\par Now the stability of a $N$-rosette will be numerically investigated. Starting from the previous numerical test we perform a dynamic simulation perturbing the initial equilibrium configuration of the $8$-rosette. Namely, each center of a follower cell is translated of a ray vector whose magnitude and direction are random number in the interval $[0,5]\;(\mu\text{m})$ and $[0,2\pi]$. Spatial domain and parameters are the same as in the previous simulation, while the time range is set to $[0,60]\;(\text{h})$. 
\par
\begin{figure}[!ht]
\centering
\subfigure[]{
\includegraphics[width=0.4\textwidth]{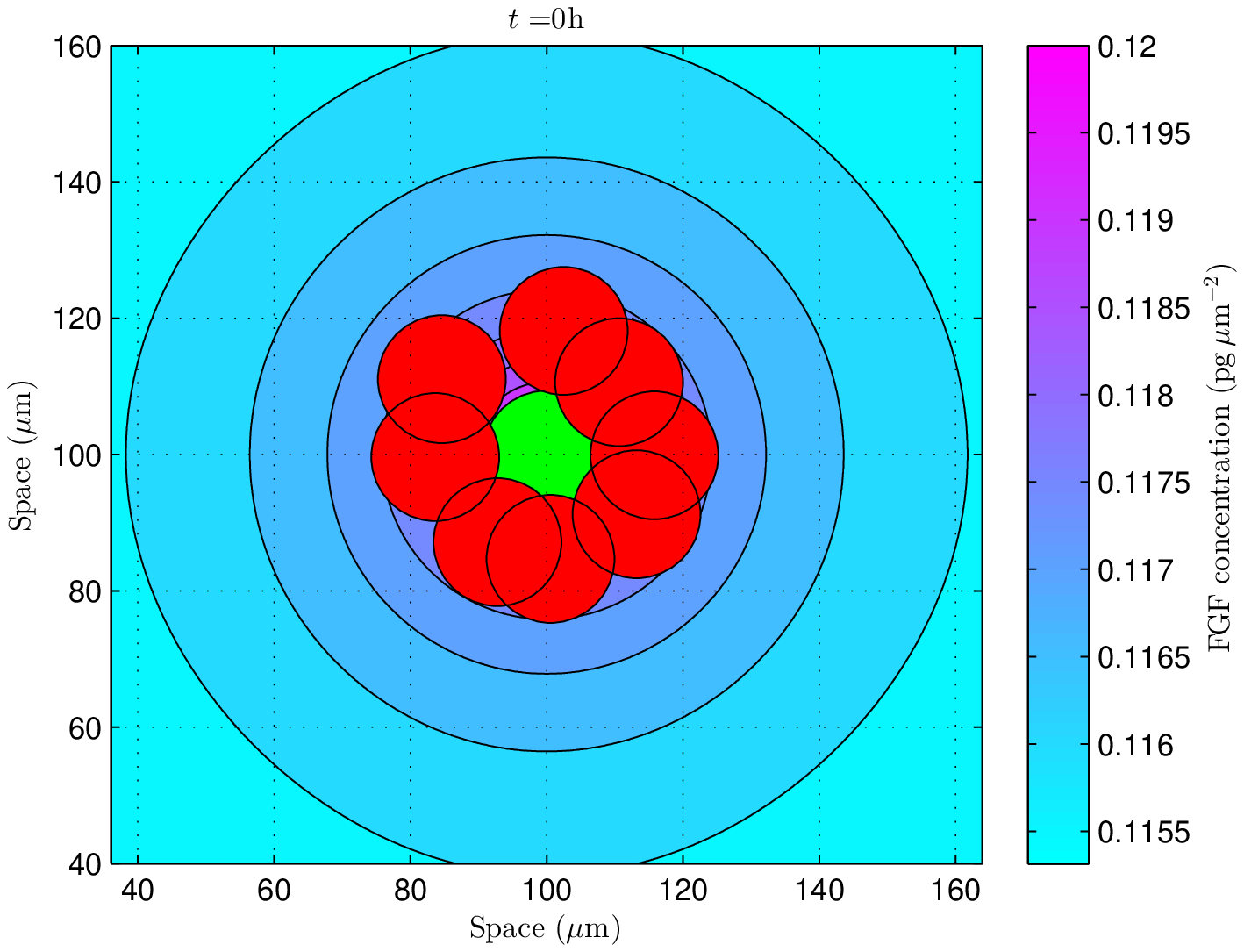}}
\hspace{0 mm}
%\subfigure[]{
%\includegraphics[width=0.4\textwidth]{stab_t20.eps}}\\
%\hspace{0 mm}
%\subfigure[]{
%\includegraphics[width=0.4\textwidth]{stab_t40.eps}}
%\hspace{0 mm}
\subfigure[]{
\includegraphics[width=0.4\textwidth]{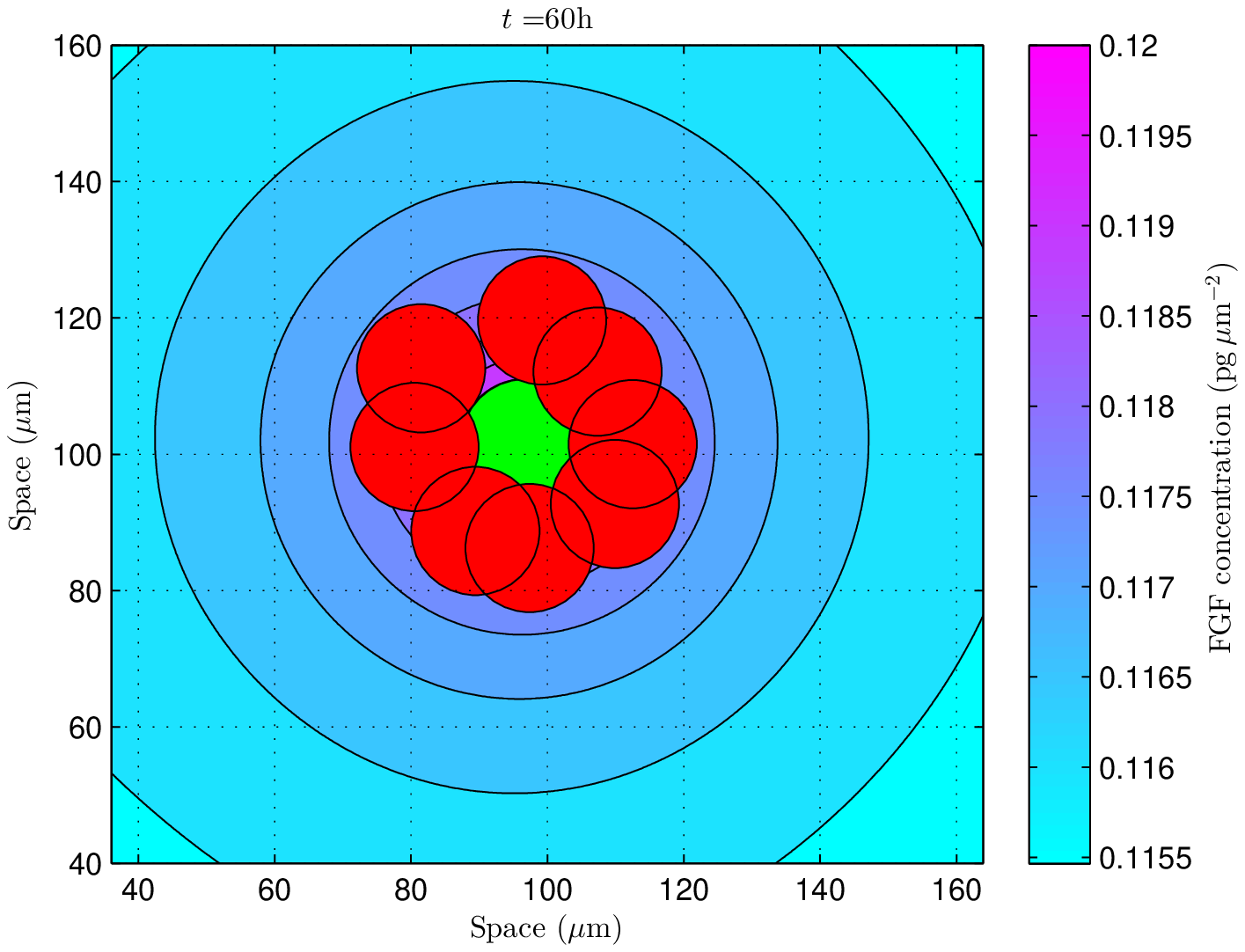}}
\caption{Numerical assessment of stability of a $8$-rosette. (a)-(b) are related respectively to the dimensional plot at $t=0\,\text{h}$ and $t=60\,\text{h}$. System (\ref{sys-adim}) is solved as described in Section \ref{ap:scheme} in $\Omega=[0,200]\times[0,200]\;(\mu\text{m}^2)$ (plot refers only to a part of the domain) and in $[0,50]\;(\text{h})$. Spatial and temporal discretization is the same as in Figure (\ref{fig:staz}). Initial data are given by a perturbation of positions (\ref{X0staz}), and by (\ref{Xdot0staz})--(\ref{s0staz}). The parameters used here are listed in Appendix \ref{ap:parameter}. Green colour ({\large{\color{green}$\bullet$}}) marks the leader cell, red colour ({\large{\color{red}$\bullet$}}) a follower cell, contour plot in the background is the FGF signal concentration. Variable $s(\mathbf{x},t)$ is not shown.}
\label{fig:stab}
\end{figure}

\begin{figure}[!ht]
\centering
\subfigure[]{\includegraphics[width=0.48\textwidth]{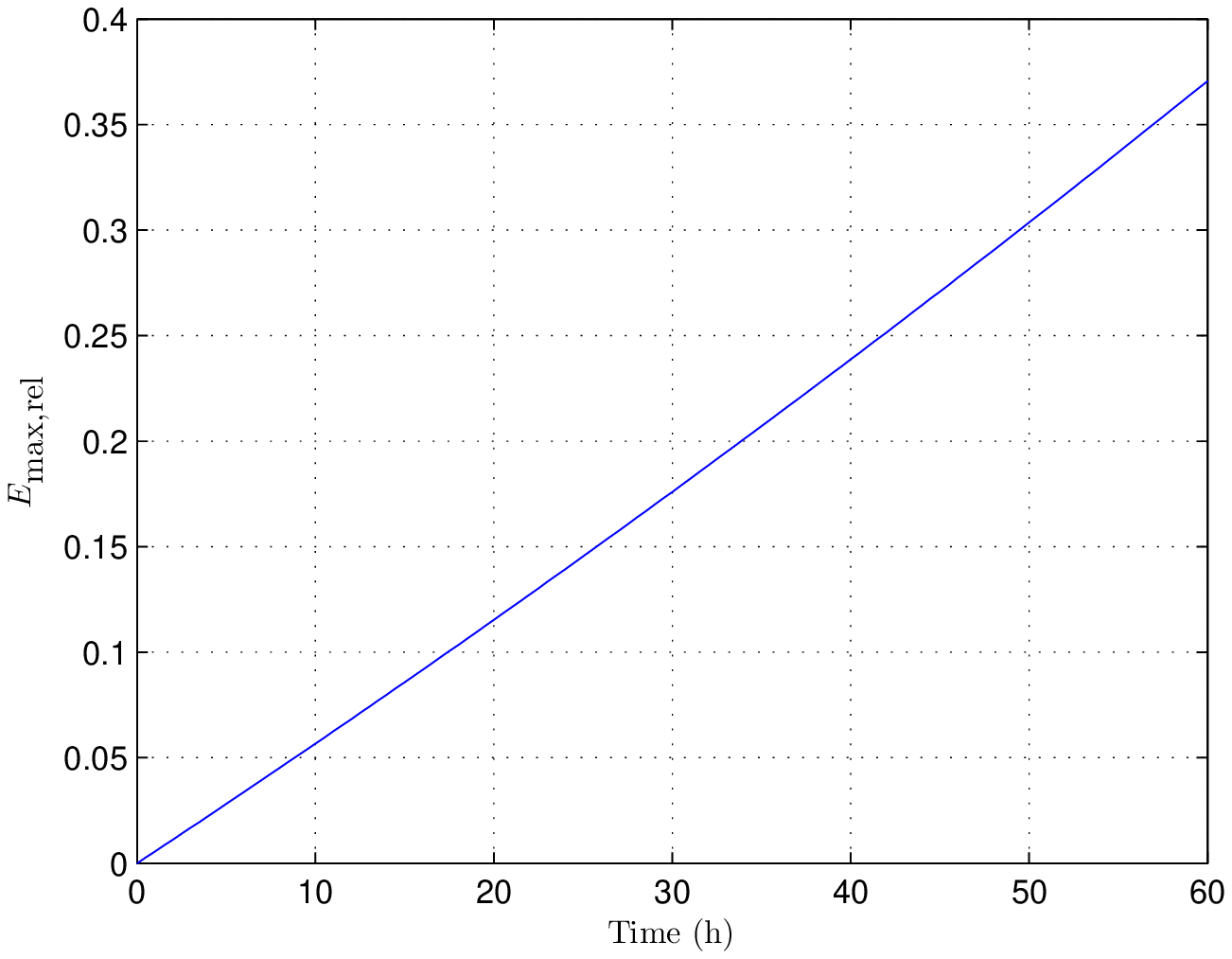}}
\hspace{1 mm}
\subfigure[]{\includegraphics[width=0.48\textwidth]{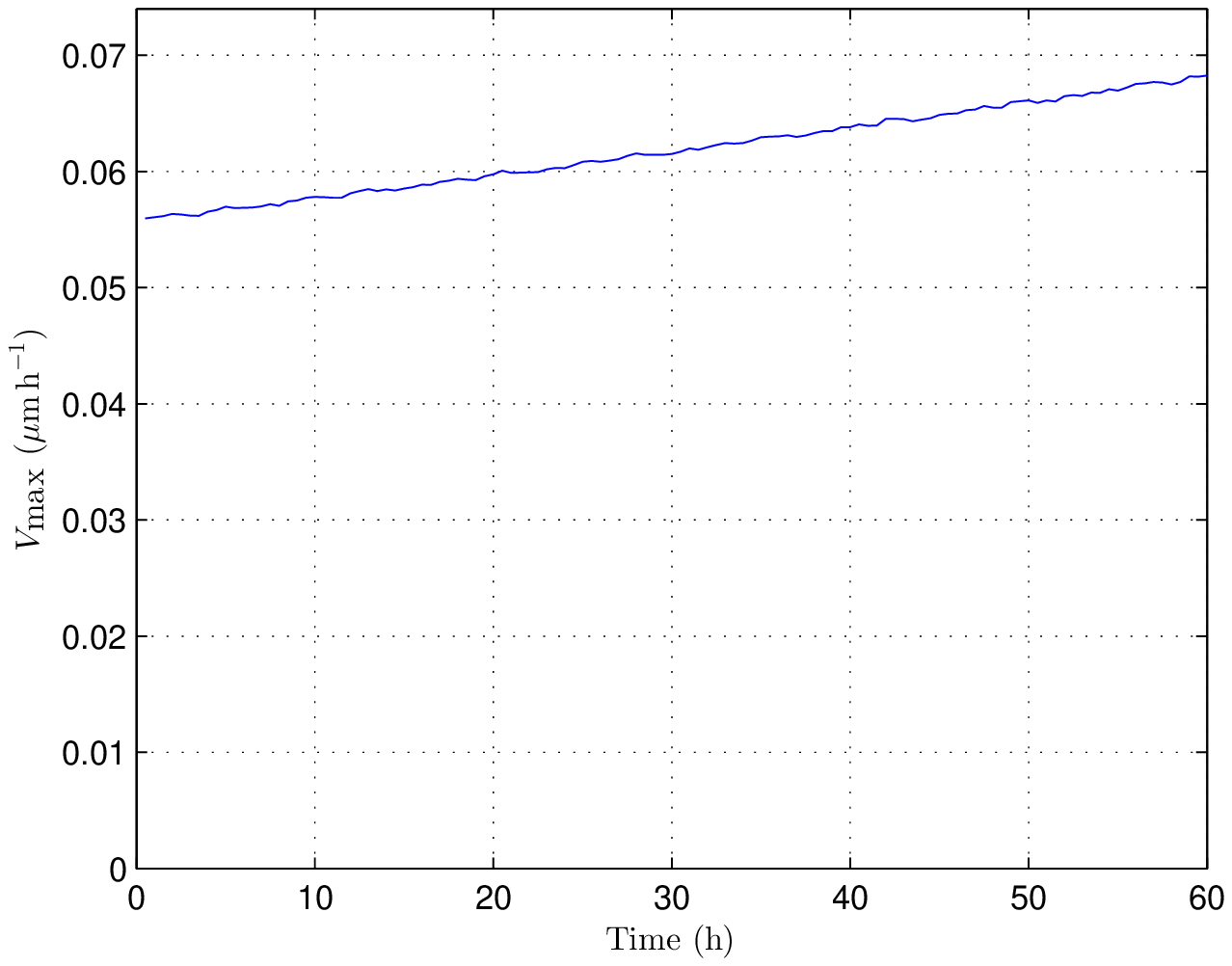}}
\hspace{1 mm}
\caption{Numerical assessment of stability of a $8$-rosette. (a) Plot in time of the maximum relative error $E_{\text{max,rel}}(t)$. (b) Plot of the maximum velocity $V_{\text{max}}(t)$.}
\label{fig:stab_err}
\end{figure}

 Figure \ref{fig:stab} shows the evolution in space of the dimensional solution at two time steps: $t=0\,\text{h}$ and $t=60\,\text{h}$. Colour convention is the same as Figure \ref{fig:stab}. Figure \ref{fig:stab_err} shows the evolution in time of the maximum relative error on the position $E_{\text{max,rel}}(t)$ and the maximum velocity $V_{\text{max}}(t)$. $E_{\text{max,rel}}$ indicates a deviation from the initial position in the order of $10^{-1}$ times cell radius, and also $V_{\text{max}}$ is small, being in the order of $10^{-2}$ $\mu\text{m}\,\text{h}^{-1}$. Our data demonstrate as the equilibrium configuration of our $8$-rosette is stable. Furthermore, numerical simulations show that similar results can hold also if $N\neq 8$, for instance for $N=5$ or $10$. We note that in a physically reasonable time range we do not observe the asymptotic stability of the rosette structures, which is actually not expected, but just the simple stability. 

\section{Dynamic simulations}\label{sec:dynamic}
\subsection{Numerical methods}\label{ap:scheme}
All the numerical tests in the paper employ a 2D finite difference scheme with a uniform spatial and temporal grid.
\par About system (\ref{sys-adim}), the equation for $\ddot{\mathbf{X}}$ is reduced to the first order system
\small
\begin{equation}\label{first-order-system}
\begin{split}
\left\{
\begin{array}{ll}
	\dot{\mathbf{Y}_i}&=\displaystyle\frac{\alpha^*}{W}\int_{\mathbf{B}(\mathbf{X}_i,\bar{R}^*)}\nabla s(\mathbf{x},t)w_i(\mathbf{x})\,d\mathbf{x}+\displaystyle	\frac{\gamma^*(1-\varphi_i)}{W}\int_{\mathbf{B}(\mathbf{X}_i,\bar{R}^*)}\nabla f(\mathbf{x},t)w_i(\mathbf{x})\,d\mathbf{x}\\\\
	&+\displaystyle\frac{1}{\bar{N}_i}\sum_{j:\mathbf{X}_j\in \mathbf{B}(\mathbf{X}_i,R^*_1)\backslash\left\{\mathbf{X}_i\right\}}\mathbf{H}(\mathbf{Y}_j-\mathbf{Y}_i)+\sum_{j:\mathbf{X}_j\in \mathbf{B}(\mathbf{X}_i,R^*_5)\backslash\left\{\mathbf{X}_i\right\}}\mathbf{K}(\mathbf{X}_j-\mathbf{X}_i)-\left[\mu^*_\text{F}+(\mu^*_\text{L}-\mu^*_\text{F})\varphi_i\right]\mathbf{Y}_i,\\\\
	\dot{\mathbf{X}_i}&=\mathbf{Y}_i.
\end{array}
	\right.
\end{split}
\end{equation}
\normalsize
Then equation (\ref{first-order-system})$_1$ is discretized with the backward Euler method, putting totally implicit the terms in $\mathbf{Y}_i$ and $\mathbf{Y}_j$ at the right hand side, while totally explicit the other addends. Equation (\ref{first-order-system})$_2$ is solved with the forward Euler method. 
\par About equation (\ref{sys-adim})$_3$ we use a classical exponential transformation in order to eliminate the stiff term $-\eta^* f$, and then we apply a central difference scheme in space and the parabolic Crank-Nicolson scheme in time, subject to zero flux boundary conditions. In practice, in the numerical simulations we choose the domain size sufficiently large that over the time period of interest have a negligible impact on the solution. 
\par Finally in equation (\ref{sys-adim})$_4$ the explicit Euler method is employed.
\subsection{Numerical tests}
Now we simulate the zebrafish lateral line growth in a two-dimensional space, during about $20\;\text{h}$. Using the numerical method proposed in Section \ref{ap:scheme} we solve system (\ref{sys-adim}) in a domain $\Omega=[0,5000]\times[0,1240]\;(\mu\text{m}^2)$, with a spatial and temporal discretization given respectively by $\Delta x=\Delta y=5\,\mu\text{m}$ and $\Delta t=0.001\,\text{h}$. Parameters values used here are listed in Appendix \ref{ap:parameter} (Tables \ref{tab-param-dim}, \ref{tab-par-adim}). Initial and boundary conditions are given by (\ref{pos-vel-iniz})--(\ref{s0-psi}). In particular, as initial datum $\mathbf{X}_i(0)$, we set 90 cells equally distributed in the stripe $[600,1180]\times[600,640]\;(\mu\text{m}^2)$ at a distance between their centers of $17\;\mu\text{m}$, and then randomized around their position with radius in the range $[0,3]\;(\mu\text{m})$ and angles in $[0,2\pi]$ (Figure \ref{fig:dinamico1} (a)). As initial condition $s(\mathbf{x},0)$ in equation (\ref{s0-tanh}) we fix $c_1=838\;\mu\text{m}$ (the inflection point of the $\tanh$ is about at the middle of the primordium), $c_2=200\;\mu\text{m}$, and $[\bar{a},\bar{b}]=[600,5000]\;(\mu\text{m})$. Then in (\ref{s0-psi}) and (\ref{eq:mollifier2}) we choose $l=20\;\mu\text{m}$ and $\varepsilon=10\;\mu\text{m}$.
\par Figures \ref{fig:dinamico1}, \ref{fig:dinamico2} show the numerical simulations of the lateral line evolution as described above at different time steps. As usual, green colour marks leader cells ($\varphi_i=1$), and red colour the followers ($\varphi_i=0$). Contour plot in the background to FGF signal, while variable $s(\mathbf{x},t)$ is not shown. In our simulation we can observe, in the first few hours after migration starts, the leader-to-follower transition of some cells in the trailing region of the primordium, up to about $t=6.5\;\text{h}$ when a first rosette starts detaching (Figures \ref{fig:dinamico1} (b)). This is consistent with the experimental results presented in the supplementary material in \cite{nechiporuk, gilmour}, that shows a time of about $3\text{--}6$ h for the first rosette separation. Figures \ref{fig:dinamico1} (c) shows the formation of a second rosette in the new trailing region, meanwhile in the first rosette the lateral inhibition process is completed leaving two leader cells. Then in the next time steps, until about $t=20\;\text{h}$, we observe the detachment of the other two rosettes (Figure \ref{fig:dinamico2}).   
\begin{figure}[!ht]
\centering
\subfigure[]{\includegraphics[width=0.8\textwidth]{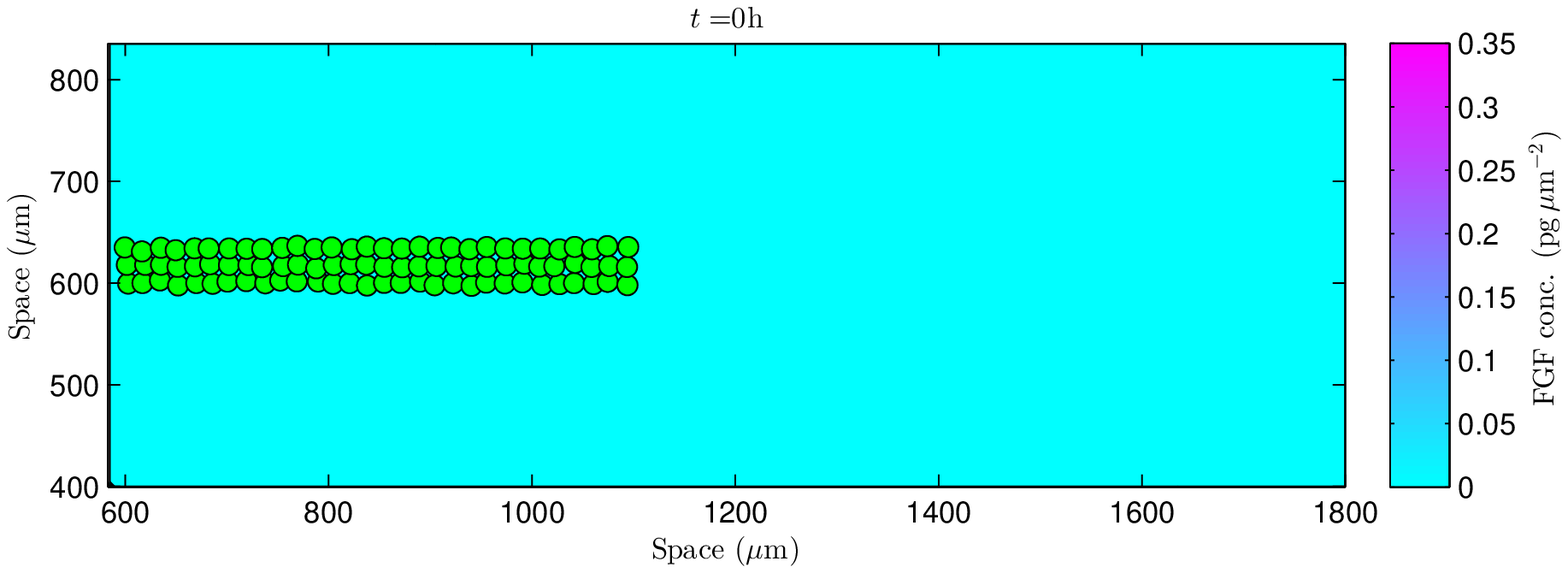}}\\
\vspace{0cm}
\subfigure[]{\includegraphics[width=0.8\textwidth]{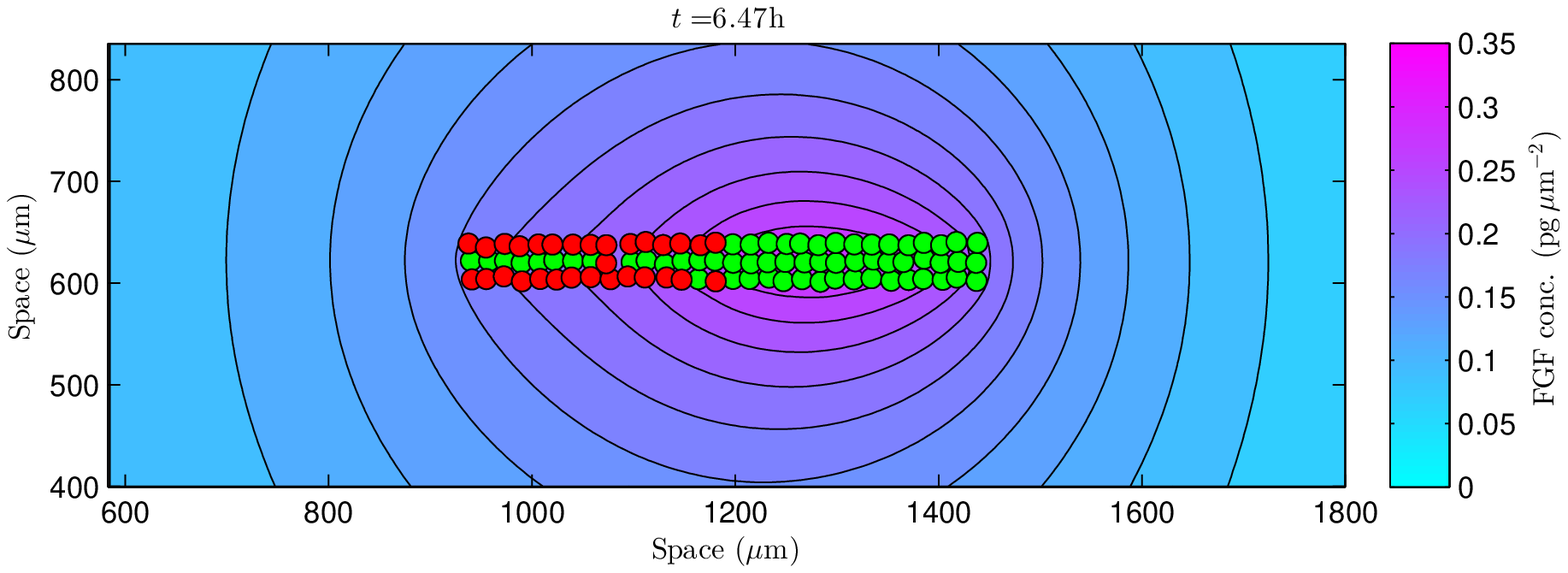}}\\
\vspace{0cm}
\subfigure[]{\includegraphics[width=0.8\textwidth]{din_t2.eps}}\\
\vspace{0cm}
\caption{Numerical simulation of the lateral line growth at five different time steps: $t=0,6.47,10.15\;\text{h}$, next two time steps $t=15.64,19\;\text{h}$ are plotted in Figure \ref{fig:dinamico2}). System (\ref{sys-adim}) is solved in the domain $\Omega=[0,5000]\times[0,1240]\;(\mu\text{m}^2)$ (plot shows only a part of the domain), with a spatial and temporal discretization given respectively by $\Delta x=\Delta y=5\,\mu\text{m}$ and $\Delta t=0.001\,\text{h}$. Parameters values used here are listed in Appendix \ref{ap:parameter}. Initial and boundary conditions are given by (\ref{pos-vel-iniz}), (\ref{phi-iniz}), (\ref{f-iniz-bound}), and (\ref{s-iniz2}). In particular, about initial condition $s(\mathbf{x},0)$, in equation (\ref{s0-tanh}) we have fixed $c_1=838\;\mu\text{m}$, $c_2=200\;\mu\text{m}$, and $[\bar{a},\bar{b}]=[600,5000]\;(\mu\text{m})$. Then in (\ref{s0-psi}) and (\ref{eq:mollifier2}) we have chosen $l=20\;\mu\text{m}$ and $\varepsilon=10\;\mu\text{m}$. Green colour ({\large{\color{green}$\bullet$}}) is for leader cells, red colour ({\large{\color{red}$\bullet$}}) for the followers. Contour plot in the background indicates the FGF concentration, while variable $s(\mathbf{x},t)$ is not shown.}
\label{fig:dinamico1}
\end{figure}
\begin{figure}[!ht]
%\ContinuedFloat
\centering
\subfigure[]{\includegraphics[width=0.8\textwidth]{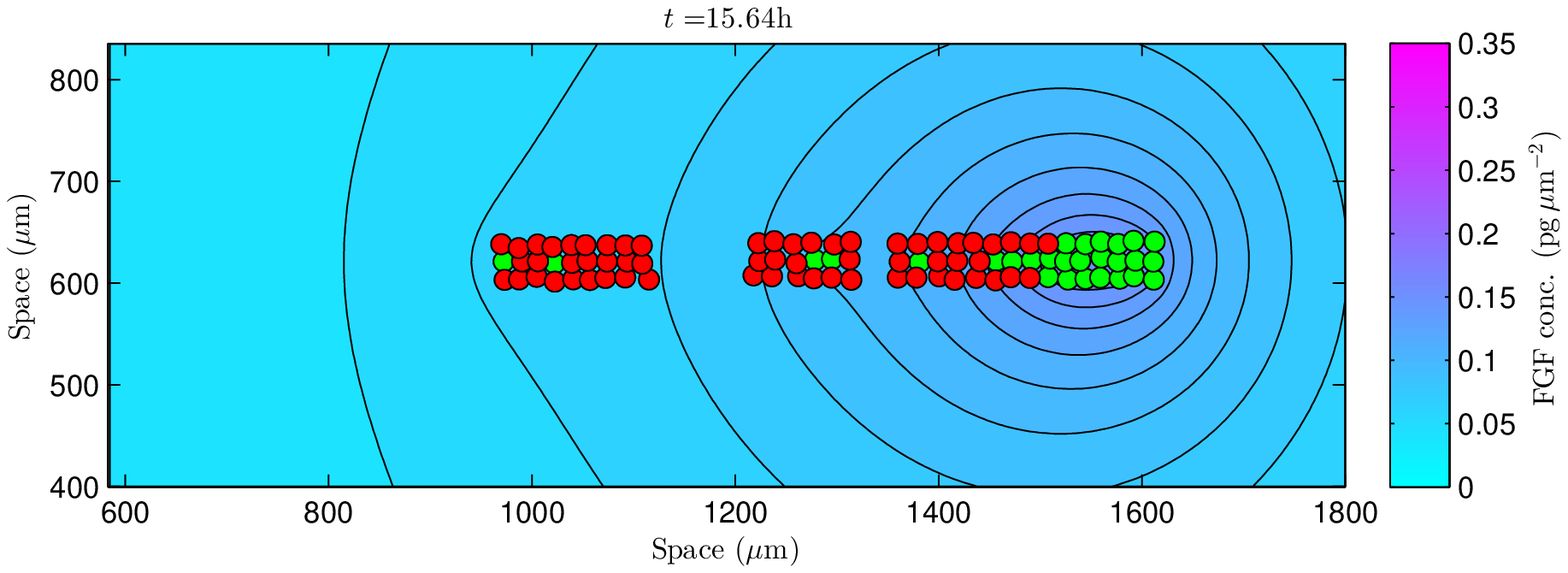}}\\
\vspace{0cm}
\subfigure[]{\includegraphics[width=0.8\textwidth]{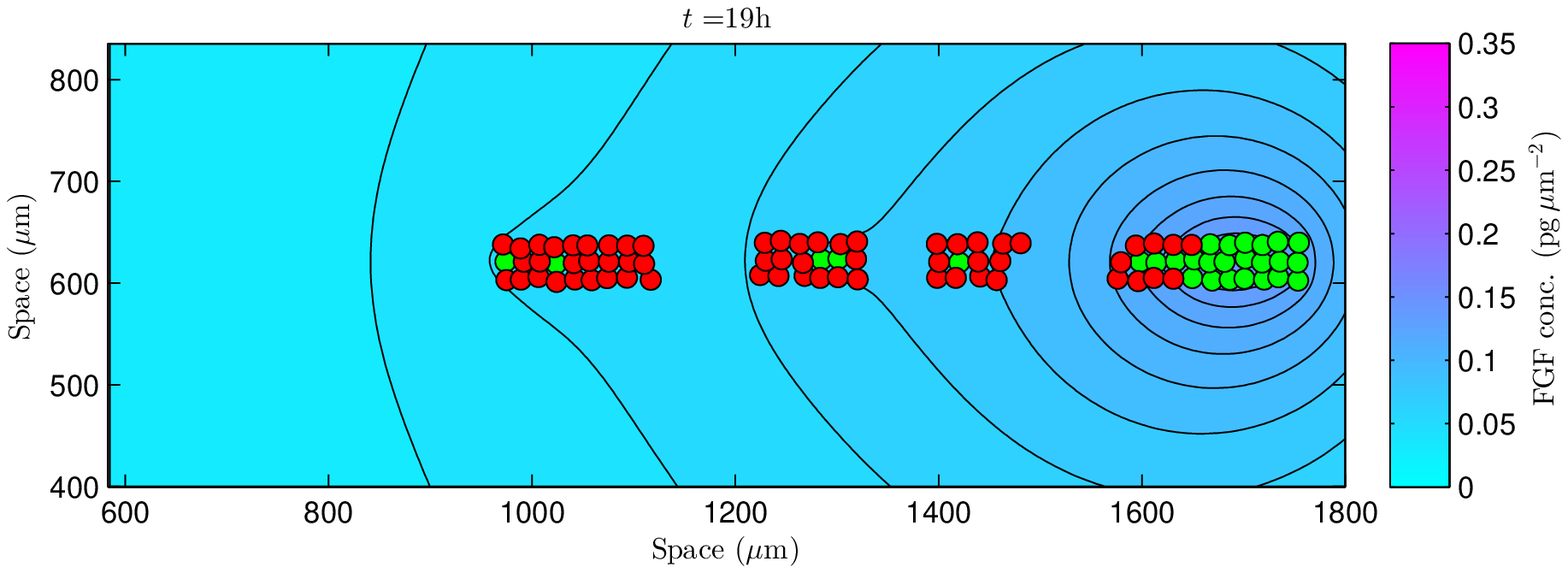}}\\
\caption[]{Continuation of Figure \ref{fig:dinamico1}. Numerical simulation of the lateral line growth at time steps: $t=15.64,19\;\text{h}$.}
\label{fig:dinamico2}
\end{figure}
\par
\begin{figure}[htbp]
	\centering
		\includegraphics[width=0.5\textwidth]{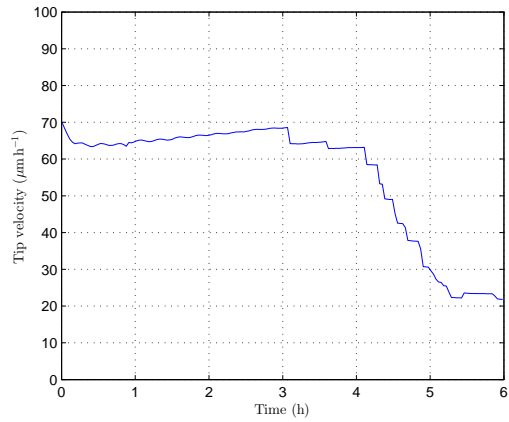}
	\caption{Numerical dimensional plot of the velocity of tip of the primordium during migration versus time.}
	\label{fig:tip_velocity}
\end{figure}

 In Figure \ref{fig:tip_velocity} we have plotted migration velocity of the tip of the primordium versus time for 6 hours. Taking into account the velocity of $69\;\mu\text{m}/\text{h}$ given in \cite{gilmour}, we can state a good concordance of our data. Moreover we observe a decrease in velocity in correspondence with the formation of the first rosette. This is substantially comparable with the velocity plot shown in \cite{gilmour} in Figure 4 (c).
\par Finally, from the numerical simulations, we observe a flocking behavior in cell migration, according to the results shown in \cite{cucker, ha} for Cucker-Smale term (\ref{alignment-new}), although in our model other effects are involved, as chemotaxis and adhesion-repulsion terms. We recall that in \cite{cucker} flocking behaviour occurs unconditionally when the power of the denominator in (\ref{alignment2}) is less than 1/2, and conditionally if this power is equal or greater than 1/2. If we consider only the equations (\ref{alignment-new})--(\ref{alignment2}) we are in the case of conditional flocking, and the flocking behaviour is ensured by the initial data (\ref{pos-vel-iniz}).  

\clearpage
\section{Conclusions}\label{sec:conclusion}
We have proposed in this paper a \emph{discrete in continuous} mathematical model describing the formation of the lateral line in the zebrafish primordium. Under suitable hypothesis, we have shown that our model admits particular biologically relevant steady solutions. They corresponds to the formation of neuromasts along the two flanks of the embryo. Then their stability has been tested numerically. Finally, the dynamical model has been tested by 2D numerical simulations and the results have been compared with some experimental observations.
\par Clearly we remark that the model proposed here presents some limits. Firstly, cells are supposed to be all equally circular, so that deformation effects are neglected. On the other hand they can be partially recovered introducing influence radii. Secondly, only a limited number of biological interactions have been modeled, and this in a time range starting with the beginning of the migration of the primordium. For example, biological phenomena occurring in the next few hours post-fertilization, or in the time after the rosettes deposition have not been taken into account. However we have followed the framework of the studies \cite{nechiporuk, gilmour}, and a good concordance with the experimental data can be inferred. 
\par Finally, we remark that with respect to \cite{nechiporuk, gilmour} we have introduced other mechanisms to obtain the global migration and the neuromast formation, as lateral inhibition, alignment, and adhesion-repulsion effects. It would be interesting to have experimental evidence in this regard.
\appendix
\section{Parameters estimates}\label{ap:parameter}
About the choice of the parameters of the model, we point out that while some values can be found or estimated from the biological or modelling literature, the others have been obtained by numerical data fitting or using some relations provided by the stationary model. 
\par Tables \ref{tab-param-dim} and \ref{tab-par-adim} summarize respectively the values of the dimensional and nondimensional parameters. In the case of a range of variability for a parameter, the selected value, used in the simulations, is put in brackets. Finally, the last column in Table \ref{tab-param-dim} specifies the references for the provided data. 
\par Now we will make some comments in this regard. Firstly, cell radius $R$ is fixed to $10\;\mu\mbox{m}$ starting from the experimental data in \cite{gilmour}. Radii $\bar{R}$, $R_1$, $R_2$, are chosen to be equal to $20\;\mu\mbox{m}$, taking into account a possible effect of cell extensions. Radius $R_3$, concerning with the range of production or degradation of a chemical signaling, is set to be equal to $R$, because we think to a source or a drain defined by the dimension of a single cell. For $R_4$ and $R_5$ we fix respectively the values $20\;\mu\mbox{m}$ and $25\;\mu\mbox{m}$. First value provides a repulsion force when two cells start to be overlapped (see equation (\ref{kappa})$_1$), second values implies an adhesion force in the spatial radial range 20--25 $\mu\mbox{m}$. The values of $\alpha$, $\beta_\text{L}$, $\beta_\text{F}$, $\gamma$, $\omega_{\text{adh,F}}$, $\mu_\text{F}$, $\delta/\lambda$, and $\sigma$, are obtained by a numerical data fitting on the respective dimensionless values, in order to obtain in the simulations a cell migration velocity and a neuromasts formation consistent with the experimental results. 
\par About the information on the parameters arising from the stationary model, we refer to formulas (\ref{kf-cond}), (\ref{kl-cond}), (\ref{omega-gamma-d}), Table \ref{tab:K}, and Figures \ref{fig:K}, \ref{fig:curve-monotonia}. The first two relations give us a limitation for $k^*_\text{F}:=k_\text{F}/\lambda$ and $k^*_\text{L}:=k_\text{L}/\lambda$, while the third one provides a value of $\omega_{\text{rep}}$ when we have fixed $\gamma$ by a numerical choice. Namely, the right hand side of these equations depend on $N$ and $d^*_1$ once the other parameters are chosen. So, to obtain the values in Tables \ref{tab-param-dim}, \ref{tab-par-adim} we have fixed for an example $N=8$ and $d^*_1=3/2 R^*$. They represent reasonable values under the experimental observations in \cite{nechiporuk} and \cite{gilmour}.
\par Then a value for $\xi$ is obtained from the respective nondimensional value $\xi^*$ in order to have $f^*_{\max}=1$ setting a single leader cell in our domain. Finally other constants are estimable from data available in literature: $s_{\max}$, from \cite{kirkpatrik}; $f_{\max}$, from \cite{walshe}; $\omega_{\text{adh,L}}$, from \cite{bell}; $\mu_\text{L}$, from \cite{rubinstein}; $D$, from \cite{yeh, filion} and a phenomenological formula in \cite{he}; $\eta$ from \cite{beenken, lee-fgf}, using the FGF half-life estimates. 
\begin{longtable}{||c|m{5cm}|m{5cm}|m{2.5cm}||}
\caption{Estimates of physical parameter values.}
\label{tab-param-dim}\\
\hline
Parameter&Definition&Value or range (used value)&Source\\
\hline
\hline
$R$&\scriptsize cell radius&10 $\mu\text{m}$&\scriptsize\cite{gilmour} \\
\hline 
\rule[-2mm]{-2mm}{0.6 cm}
$\bar{R}$&\scriptsize detection radius of chemicals &20 $\mu\text{m}$&\scriptsize biological assumption\\
\hline
$R_1$&\scriptsize detection radius of cellular alignment&20 $\mu\text{m}$&\scriptsize biological assumption\\
\hline
$R_2$&\scriptsize detection radius of lateral inhibition&20 $\mu\text{m}$&\scriptsize biological assumption\\
\hline
$R_3$&\scriptsize radius of production/degradation of chemicals&10 $\mu\text{m}$&\scriptsize biological assumption\\
\hline
$R_4$&\scriptsize radius of action of repulsion between cells&20 $\mu\text{m}$&\scriptsize biological assumption\\
\hline
$R_5$&\scriptsize radius of action of adhesion between cells&25 $\mu\text{m}$&\scriptsize biological assumption\\
\hline
T&\scriptsize characteristic time&$1/3600\,\text{h}$&\scriptsize \cite{rubinstein}\\
\hline
$s_{\max}$&\scriptsize maximum concentration of SDF-1a&\parbox[t][][t]{5.5cm}{$3.6\times10^{-8}\text{--}6.5\times10^{-8}$\\$(2.5\times10^{-8})\;\text{pg}\,\mu \text{m}^{-2}$}&\scriptsize \cite{kirkpatrik}\\
\hline
$f_{\max}$&\scriptsize maximum concentration of FGF&$1\times10^{-1}\text{--}(1.2\times10^{-1})\;\text{pg}\,\mu \text{m}^{-2}$&\scriptsize \cite{walshe}\\
\hline
%$c_1$&coefficient for the initial concentration of SDF-1a& &assumed\\
%\hline
%$c_2$&coefficient for the initial concentration of SDF-1a& &assumed\\
%\hline
%$\varepsilon$&coefficient for the mollifier (\ref{eq:mollifier}) &$10\;\mu\text{m}$&assumed\\
%\hline
$\Gamma_0$&\scriptsize constant in function (\ref{gamma-n-i})&10 nondim.&\scriptsize assumed\\
\hline
$\alpha$&\scriptsize coefficient of SDF-1a haptotactic effect per unit mass&$1.31\times10^{27}\;\mu\text{m}^{4}\,\text{h}^{-2}\,\text{pg}^{-1}$&\scriptsize assumed\\
\hline
$\beta_\text{L}$&\scriptsize coefficient of cell flocking per unit mass for a leader cell&$5\times10^{20}\;\text{h}^{-1}$&\scriptsize assumed\\
\hline
$\beta_\text{F}$&\scriptsize coefficient of cell flocking per unit mass for a follower cell&$5\times10^{18}\;\text{h}^{-1}$&\scriptsize assumed\\
\hline
$\gamma$&\scriptsize coefficient of attraction toward FGF source per unit mass&$1.08\times 10^{20}\;\mu \text{m}^{4}\,\text{h}^{-2}\,\text{pg}^{-1}$&\scriptsize assumed\\
\hline
$\omega_{\text{rep}}$&\scriptsize coefficient of repulsion per unit mass&$2.03\times 10^{17}\;\mu\text{m}^{2}\,\text{h}^{-2}$&\scriptsize from steady model, formula (\ref{omega-gamma-d})\\
\hline
$\omega_{\text{adh,L}}$&\scriptsize elastic constant per unit mass for a leader cell&\parbox[t][][t]{5.5cm}{$1.296\times 10^{14}\text{--}1.296\times 10^{19}$\\$(5.5\times 10^{16})\;\text{h}^{-2}$}&\scriptsize \cite{bell}\\
\hline
$\omega_{\text{adh,F}}$&\scriptsize elastic constant per unit mass for a follower cell&\parbox[t][][t]{5.5cm}{$1.296\times 10^{12}\text{--}1.296\times 10^{17}$\\$(5.5\times 10^{14})\;\text{h}^{-2}$}&\scriptsize assumed\\
\hline
$\mu_\text{L}$&\scriptsize damping coefficient for a leader cell per unit mass&$(5.82\times 10^{14})\text{--}5.82\times 10^{15}$ $\text{h}^{-1}$&\scriptsize \cite{rubinstein}\\
\hline
$\mu_\text{F}$&\scriptsize damping coefficient for a follower cell per unit mass&$(8\times 10^{15})\text{--}8\times 10^{16}\;\text{h}^{-1}$&\scriptsize assumed\\
\hline
$\delta/\lambda$&\scriptsize ratio of coefficient of sensibility to SDF-1a and coefficient of lateral inhibition&$1.12\times 10^9$ $\;\text{pg}^{-1}\,\mu \text{m}^{2}$&\scriptsize assumed\\
\hline
$k_\text{L}/\lambda$&\scriptsize ratio of coefficient of sensibility to FGF signal for a leader cell and coefficient of lateral inhibition&$<1.8187$ (1.7) nondim.&\scriptsize from steady model, formula (\ref{kl-cond})\\
\hline
$k_\text{F}/\lambda$&\scriptsize ratio of coefficient of sensibility to FGF signal for a follower cell and coefficient of lateral inhibition&$\geq 1.1619$ (17) nondim.&\scriptsize from steady model, formula (\ref{kf-cond})\\
\hline
%$\lambda$&coefficient of sensibility to the lateral inhibition&$\;\text{pg}\,\mu \text{m}^{-2}$&from nondimensional parameters\\
%\hline
$D$&\scriptsize diffusion coefficient&69985--84184 (78950) $\mu \text{m}^2\,\text{s}^{-1}$&\scriptsize \cite{he, filion, yeh}\\
\hline
$\xi$&\scriptsize coefficient of production of FGF&$2.9592\;\text{pg}\,\mu \text{m}^{-2}\,\text{h}^{-1}$&\scriptsize assumed\\
\hline
$\eta$&\scriptsize degradation constant of FGF&0.09--0.69 (0.2) $\text{h}^{-1}$\; &\scriptsize \cite{beenken, lee-fgf}\\
\hline
$\sigma$&\scriptsize degradation constant of SDF-1a&%0.096\mbox{--}1.61
0.6\;$\text{h}^{-1}$&\scriptsize assumed\\
\hline
\hline
\end{longtable}
\newpage
\begin{longtable}{||c|c|m{4.5cm}||}
\caption{Estimates of dimensionless parameter values.}
\label{tab-par-adim}\\
\hline
Parameter&Definition&Value or range (used value)\\
\hline
\hline
%\rule[-2mm]{-2mm}{0.6 cm}
%$R^*$&--&1\\
%\hline 
\rule[-2mm]{-2mm}{0.6 cm}
$\bar{R}^*$&$\bar{R}/R$&2 \\
\hline
\rule[-2mm]{-2mm}{0.6 cm}
$R^*_1$&$R_1/R$&2\\
\hline
\rule[-2mm]{-2mm}{0.6 cm}
$R^*_2$&$R_2/R$&2\\
\hline
\rule[-2mm]{-2mm}{0.6 cm}
$R^*_3$&$R_3/R$&1\\
\hline
\rule[-2mm]{-2mm}{0.6 cm}
$R^*_4$&$R_4/R$&2\\
\hline
\rule[-2mm]{-2mm}{0.6 cm}
$R^*_5$&$R_5/R$&2,5\\
\hline
%\rule[-2mm]{-2mm}{0.6 cm}
%$s^*_{\max}$&--&1\\
%\hline
%\rule[-2mm]{-2mm}{0.6 cm}
%$f^*_{\max}$&--&1\\
%\hline
\rule[-2mm]{-2mm}{0.6 cm}
%$c^*_1$&&from dimensional parameters\\
%\hline
%$c^*_2$&&from dimensional parameters\\
%\hline
%$\varepsilon^*$&1&from dimensional parameters\\
%\hline
$\Gamma_0$&$\Gamma_0$&10\\
\hline
\rule[-2mm]{-2mm}{0.6 cm}
$\alpha^*$&$\alpha s_{\max} T^2/R^2$   &$2.53\times 10^{10}$\\
\hline
\rule[-2mm]{-2mm}{0.6 cm}
$\beta_\text{L}^*$&$\beta_\text{L} T$&$1.39\times 10^{17}$\\
\hline
\rule[-2mm]{-2mm}{0.6 cm}
$\beta_\text{F}^*$&$\beta_\text{F} T$&$1.39\times 10^{15}$\\
\hline
\rule[-2mm]{-2mm}{0.6 cm}
$\gamma^*$&$\gamma f_{\max} T^2/R^2$&$10^{10}$\\
\hline
\rule[-2mm]{-2mm}{0.6 cm}
$\omega^*_{\text{rep}}$&$\omega_{\text{rep}} T^2/R^2$&$1.57\times 10^{8}$\\
\hline
\rule[-2mm]{-2mm}{0.6 cm}
$\omega^*_{\text{adh,L}}$&$\omega_{\text{adh,L}}T^2$&$10^7$--$10^{12}$ ($4.24\times 10^{9}$)\\
\hline
\rule[-2mm]{-2mm}{0.6 cm}
$\omega^*_{\text{adh,F}}$&$\omega_{\text{adh,F}}T^2$&$10^5$--$10^{10}$ ($4.24\times 10^{7}$)\\
\hline
\rule[-2mm]{-2mm}{0.6 cm}
$\mu^*_\text{L}$&$\mu_\text{L} T$&$(1.62\times 10^{11})$--$1.62\times 10^{12}$\\
\hline
\rule[-2mm]{-2mm}{0.6 cm}
$\mu^*_\text{F}$&$\mu_\text{F} T$&$(2.22\times 10^{12}$)--$2.21\times 10^{13}$\\
\hline
\rule[-2mm]{-2mm}{0.6 cm}
$\delta^*$&$\delta s_{\max}/\lambda$&28\\
\hline
\rule[-2mm]{-2mm}{0.6 cm}
$k^*_\text{L}$&$k_\text{L}/\lambda$&$<1.8187$ (1.7)\\
\hline
\rule[-2mm]{-2mm}{0.6 cm}
$k^*_\text{F}$&$k_\text{F}/\lambda$&$\geq 1.1619$ (17)\\
\hline
\rule[-2mm]{-2mm}{0.6 cm}
$D^*$&$D T/R^2$&0.1944--0.2338 (0.2193)\\
\hline
\rule[-2mm]{-2mm}{0.6 cm}
$\xi^*$&$\xi T/f_{\max}$&0.0069\\
\hline
\rule[-2mm]{-2mm}{0.6 cm}
$\eta^*$&$\eta T$&\parbox[t][][t]{4.5cm}{$2.5\times 10^{-5}$--$1.92\times 10^{-4}$ \\($5.56\times 10^{-5}$)}\\
\hline
\rule[-2mm]{-2mm}{0.6 cm}
$\sigma^*$&$\sigma T$&$1.67\times 10^{-4}$\\
\hline
\hline
\end{longtable}
%
%\begin{acknowledgements}
\bigskip
\subsection*{Acknowledgements} We thank Andrea Tosin for some useful discussions and suggestions. The research leading to these results has received funding from the
European Union Seventh Framework Programme [FP7/2007-2013]  under grant agreement n. 257462 HYCON2 Network of excellence. This work has also been partially supported by the PRIN project 2008-2009 ``Equazioni iperboliche non lineari e fluidodinamica''.
%\end{acknowledgements}
%
%

\end{document}